\documentclass[aps,pra,twocolumn,superscriptaddress]{revtex4-2}

\usepackage{mathtools}
\usepackage{amssymb}
\usepackage{graphicx}
\usepackage{bbold}
\usepackage{hyperref}
\usepackage{xcolor}
\usepackage{enumerate}

\hypersetup{colorlinks=true, linkcolor=blue, urlcolor=blue, citecolor=blue}

\newcommand{\sys}{\mathcal S}
\newcommand{\env}{\mathcal E}
\newcommand{\uba}{Departamento de F\'\i sica, FCEyN, UBA, Pabell\'on 1, Ciudad Universitaria, 1428 Buenos Aires, Argentina}
\newcommand{\ifiba}{Instituto de F\'\i sica de Buenos Aires, UBA CONICET, Pabell\'on 1, Ciudad Universitaria, 1428 Buenos Aires, Argentina}

\begin{document}
\title{General theory for thermal and nonthermal quantum linear engines}

\author{Milton Aguilar}
	\email{mil@df.uba.ar}
	\affiliation{\uba}
	\affiliation{\ifiba}

\author{Juan Pablo Paz}
	\email{paz@df.uba.ar}
	\affiliation{\uba}
	\affiliation{\ifiba}

\date{\today}

\begin{abstract}
We present the exact theory of quantum engines whose working medium is a network of driven oscillators performing an arbitrary cyclic process while coupled to thermal and nonthermal reservoirs. We show that when coupled to a single reservoir work cannot be extracted unless there is population inversion, and prove that the ratio between the heat flowing out and into the working medium cannot be arbitrarily small, satisfying a form of Clausius inequality. We use such identity to prove that the efficiency of linear quantum engines satisfies a generalized bound, which coincides with the Carnot limit for thermal reservoirs. The previous results enable us to estimate the cost of preparing nonthermal reservoirs, which, if available, could be used to violate the Carnot limit.
\end{abstract}

\maketitle

\section{Introduction}
The development of thermodynamics has been centered around heat engines: machines that run cyclically between thermal reservoirs and convert heat into work. The advent of quantum thermodynamics \cite{andersVinjanampathy,gooldHuberRiera,talknerHanggi} has opened the door to a new class of engines, where the working medium and the reservoirs behave quantum mechanically \cite{gevaKosloff,scullySZAgarwal,quanLiuSun,GKAlickiKurizki,solinasAverinPekola,abahLutz,kosloffLevy,skSingerLutz,elouardHMHuard,leitchPiccioneBellomo}. Although many examples have been analyzed, no exact general theory for quantum engines has been developed (there are, however, general theories developed under various assumptions \cite{alicki,gevaKosloff2,quanZhangSun,alickiGK,sonTalknerThigna}). In this work we present such a theory for engines performing an arbitrary cyclic process with a working medium composed of driven oscillators, and coupled to thermal and nonthermal reservoirs. We show, from first principles, that these engines satisfy a form of Clausius inequality and also that their efficiency is bounded by a generalized bound, which coincides with the Carnot limit when the engine is coupled to thermal reservoirs. Finally, we estimate the energetic cost of preparing nonthermal reservoirs (which enable the engine to reach efficiencies greater than the Carnot one) from thermal ones, and relate it to the energy needed to run the engine and the work produced by it. For the sake of reading clarity we will postpone the detailed description of our model until the very end. 

The paper is organized as follows. In Sec. \ref{sec:engine} we describe the basic nature of quantum linear engines, including how to perform any thermodynamic cycle. In Sec. \ref{sec:workHeat} we show how to compute the work produced by the engine and the heat exchanged with the reservoirs. In Sec. \ref{sec:averageWork} we explicitly compute the average work produced in a cycle using the heat exchanged by the reservoirs. In Sec. \ref{sec:planck} we use our previous results to derive a generalized version of the second law of thermodynamics. In Sec. \ref{sec:clausius} we find a simple bound for the ratio between the heat flowing out and into the working medium in the form of Clausius inequality. In Sec. \ref{sec:efficiency} we use the previous inequality to prove the efficiency of these engines satisfy a generalized bound. In Sec. \ref{sec:cost} we show how to estimate the cost of preparing nonthermal reservoirs that could be used to achieve efficiencies greater than the classical Carnot limit. In Sec. \ref{sec:model} we present the description of our model. Finally, we summarize our results in Sec. \ref{sec:conclusions}.

\section{The engine \label{sec:engine}}
All engines are composed of two basic parts: a working medium that transforms heat into work, and a collection of reservoirs that act as sources or sinks for that heat. The working medium of a linear quantum engine is a network of oscillators $\sys$ with variable frequencies and couplings (see Ref. \cite{martinezPaz} for the static case). Coupled to $\sys$ there is an environment $\env$, which is composed of different pieces $\env_{\alpha}$ that will play the role of reservoirs. Each $\env_{\alpha}$ is formed by a collection of noninteracting harmonic oscillators, and it is initially prepared in an arbitrary uncorrelated state. The evolution of the engine is governed by the Hamiltonian $H = H_{\sys} (t) + H_{\env} + H_{\sys , \env}$. The engine begins its operation with its working medium and reservoirs uncorrelated but this rapidly changes due to the presence of the interaction Hamiltonian $H_{\sys , \env}$, which is assumed to be bilinear in the position coordinates of all the components of $\sys$ and $\env$.

\begin{figure}[htp]
    \centering
    \includegraphics[scale=.76]{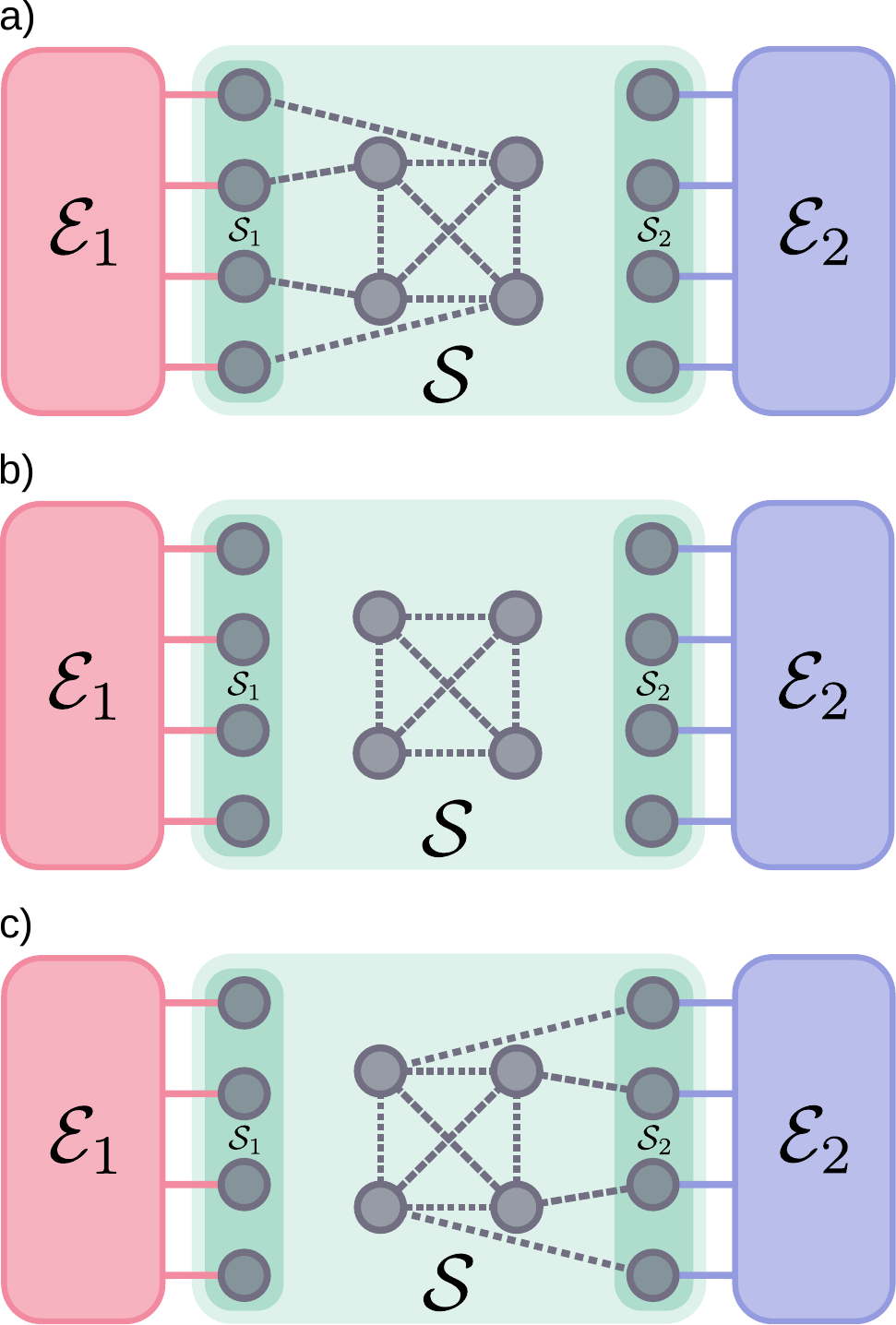}
    \caption{An engine using two reservoirs $\env_{1}$ and $\env_{2}$. $\sys$ is composed of 12 oscillators, four of which belong to $\sys_{1}$ and another four to $\sys_{2}$. $\sys_{1}$ and $\sys_{2}$ are always coupled to $\env_{1}$ and $\env_{2}$ (solid lines), respectively, and temporarily coupled to the rest of $\sys$ (dashed lines). In this setup, $\sys_{\bar{1}}$ are the four center oscillators plus the ones in $\sys_{2}$. Analogously, $\sys_{\bar{2}}$ are the four center oscillators plus the ones in $\sys_{1}$. (a) $\sys$ is only coupled to $\env_{1}$. (b) $\sys$ is decoupled from $\env_{1}$ and $\env_{2}$. (c) $\sys$ is only coupled to $\env_{2}$.}
    \label{fig:cycle}
\end{figure}

Any thermodynamic cycle performed by a linear quantum engine can be described by our model. A cycle is composed by a sequence of processes, each of which either changes some property of the working medium $\sys$, or couples or decouples it with a reservoir $\env_{\alpha}$. The first kind of processes can be realized by changing $H_{\sys}$, which has the form $H_{\sys} (t) = [P^{T} M^{-1} P + X^{T} V(t) X] / 2$ ($X$ and $P$ are vectors that contain the coordinates and momenta of the components of $\sys$, respectively, and $M$ and $V(t)$ are real matrices). By varying $V(t)$, which is assumed to be $\tau_{d}$ periodic, one changes the frequencies of the oscillators of $\sys$ as well as the interactions between them. In contrast, the coupling and decoupling with the reservoirs is done in two steps. First, for each $\env_{\alpha}$ a portion $\sys_{\alpha}$ of the oscillators of $\sys$ are selected to be permanently coupled to $\env_{\alpha}$. These oscillators are highly overdamped so as to closely follow the state of $\env_{\alpha}$, and will act as bridges between $\env_{\alpha}$ and the rest of $\sys$, which we denote $\sys_{\bar{\alpha}}$ (i.e., $\sys_{\bar{\alpha}} = \sys \setminus \sys_{\alpha}$). And second, the couplings between $\sys_{\alpha}$ and $\sys_{\bar{\alpha}}$, which are contained in $V(t)$, are modified in such a way that they are either zero or nonzero for some required time. In this way the interaction between $\sys_{\bar{\alpha}}$ and $\env_{\alpha}$ can be turned on and off and, thus, the working medium can be effectively uncoupled from $\env_{\alpha}$. By changing the strength of the interaction between $\sys_{\alpha}$ and $\sys_{\bar{\alpha}}$, both equilibrium and nonequilibrium processes can be performed. See Fig. \ref{fig:cycle} for an illustration of this scheme.

\section{Work and heat \label{sec:workHeat}}
The power produced and the heat exchanged by $\sys$ are defined through Heisenberg's equation for $H_{\sys}$:
\begin{equation}
    d \langle H_{\sys} \rangle / dt = - i \langle  [ H_{\sys} , H_{\sys , \env} ] \rangle / \hbar + \langle \partial H_{\sys} / \partial t \rangle.
    \label{variationHS}
\end{equation}
The first term in Eq. \eqref{variationHS} represents the change of the energy of $\sys$ due to the interaction with $\env$, and is associated with the heat current $\dot{Q}_{\sys}$. The second term in Eq. \eqref{variationHS} accounts for the variation of the energy of $\sys$ due to the explicit time dependence of $H_{\sys}$, and is associated with the power $\dot{W}_{\sys}$ injected into $\sys$ by the driving field. Using these definitions Eq. \eqref{variationHS} reads as: $ d \langle H_{\sys} \rangle / dt = \dot{Q}_{\sys} + \dot{W}_{\sys}$. In Ref. \cite{freitasPaz17} it was shown that the coupling with the reservoirs can induce a stationary regime where the state of $\sys$ is $\tau_{d}$ periodic. This has two consequences: (i) the average of $d \langle H_{\sys} \rangle / dt$ over $\tau_{d}$ vanishes; and (ii) the average heat current $\dot{\bar{Q}}_{\sys}$ is determined by the average variation of the energy of the reservoirs (i.e., $\dot{\bar{Q}}_{\sys} = - \dot{\bar{Q}}_{\env}$, where $\dot{\bar{Q}}_{\env}$ is the average of $d \langle H_{\env} \rangle / dt $. See Appendix \ref{sec:heatCurrents} for the proof). Thus, the time average of Eq. \eqref{variationHS} implies that $0 = \dot{\bar{Q}}_{\env} - \dot{\bar{W}}_{\sys}$. Since, as shown in Appendix \ref{sec:energy}, $\dot{\bar{Q}}_{\env}$ is constant in the stationary regime, then $\dot{\bar{Q}}_{\env} = \Delta \mathcal{Q} / \tau_{d}$, where $\Delta \mathcal{Q}$ is the average heat exchanged by the reservoirs in one period of the driving. Therefore, from Eq. \eqref{variationHS} we find that the average work performed on $\sys$ during a cycle is such that $\bar{W}_{\sys} = \Delta \mathcal{Q}$.

\section{Average work in a cycle \label{sec:averageWork}}
Our results are a consequence of three key properties that generalize the ones analyzed for thermal reservoirs in Refs. \cite{freitasPaz17,freitasPaz18,aguilarFreitasPaz,aguilarPaz}. First, in the stationary regime $\Delta \mathcal{Q}$ depends only on one property of the initial state of the reservoirs: the expectation value of the number operator $n_{\alpha} (\omega) = \langle a_{\alpha , \omega}^{\dagger} a_{\alpha , \omega} \rangle$ ($a_{\alpha , \omega}^{\dagger}$ and $a_{\alpha , \omega}$ are the creation and annihilation operators of the mode with frequency $\omega$ in $\env_{\alpha}$). In fact, as shown in Appendix \ref{sec:energy}, all other moments of $a_{\alpha , \omega}^{\dagger}$ and $a_{\alpha , \omega}$ do not contribute to time-extensive terms in $\langle H_{\env} \rangle$. Second, $\sys$ and $\env$ only exchange packets of energy with values $k \hbar \omega_{d}$ ($k$ denotes a positive integer and $\omega_{d} = 2 \pi / \tau_{d}$ is the driving frequency). Third, only two processes are relevant for this energy exchange and, as a consequence, $\Delta \mathcal{Q}$ can be written as the sum of two terms: $\Delta \mathcal{Q} = \Delta \mathcal{Q}^{NR} + \Delta \mathcal{Q}^{R}$. The explicit form of $\Delta \mathcal{Q}^{NR}$ and $\Delta \mathcal{Q}^{R}$ is derived from first principles in Appendix \ref{sec:energy}. $\Delta \mathcal{Q}^{NR}$, which is always positive, carries the effect of the nonresonant processes that transform a packet of energy $k \hbar \omega_{d}$ of the driving field into excitations of two environmental modes with frequencies $\omega_{i}$ and $\omega_{j}$ such that $\omega_{i} + \omega_{j} = k \omega_{d}$. It can be written as
\begin{equation}
    \Delta \mathcal{Q}^{NR} = \tau_{d} \sum_{k, \alpha , \beta} k \hbar \omega_{d} \int_{0}^{k \omega_{d}} d \omega \, \tilde{p}_{\alpha \beta}^{(k)} (\omega) [ n_{\beta} (\omega) + 1/2],
    \label{qNR}
\end{equation}
where $\tilde{p}_{\alpha \beta}^{(k)}$ is a positive dimensionless quantity that can be interpreted as an emission rate per unit frequency from the driving field into environmental excitations, one in the mode with frequency $k \omega_{d} - \omega$ in $\env_{\alpha}$ and another one in the one with frequency $\omega$ in $\env_{\beta}$. $\Delta \mathcal{Q}^{R}$ carries the effect of the resonant processes, which are responsible for the transport of excitations between different environmental modes due to the absorption (or emission) of a packet of energy $k \hbar \omega_{d}$ from (or into) the driving field. It reads
\begin{equation}
    \Delta \mathcal{Q}^{R} = \tau_{d} \sum_{k, \alpha , \beta} k \hbar \omega_{d} \int_{0}^{\infty} d \omega \, p_{\alpha \beta}^{(k)} (\omega) [n_{\beta} (\omega) - n_{\alpha} (\omega + k \omega_{d})].
    \label{qR}
\end{equation}
Here, $p_{\alpha \beta}^{(k)}$ is a positive dimensionless quantity that can be interpreted as a transition rate per unit frequency between the mode with frequency $\omega$ in $\env_{\beta}$ and the one with frequency $\omega + k \omega_{d}$ in $\env_{\alpha}$ (which, as shown in Appendix \ref{sec:energy}, is identical to the transition rate of the reverse process). Notably, the integrands in $\Delta \mathcal{Q}^{NR}$ and $\Delta \mathcal{Q}^{R}$ are a product of a term that depends on the cycle the engine is performing and the spectral properties of the reservoirs ($\tilde{p}_{\alpha \beta}^{(k)}$ or $p_{\alpha \beta}^{(k)}$), and a term depending only on the initial occupation number $n_{\alpha}$ of the reservoirs.

\section{Planck's proposition \label{sec:planck}}
The second law of thermodynamics can be simply obtained from our previous equations: if all the reservoirs are initially prepared in states with the same occupation number $n_{\alpha} (\omega)$, and $n_{\alpha} (\omega)$ is a decreasing function of $\omega$, then $\Delta \mathcal{Q}^{R} > 0$ and, thus, $\bar{W}_{\sys}$ is positive too. This is a generalization of Planck's proposition that affirms it is impossible to extract work from an engine working cyclically while coupled single thermal reservoir \cite{planck}. Instead, if there is some population inversion (i.e., if $n_{\alpha} (\omega)$ is not a decreasing function of $\omega$), then the reservoir is thermodynamically unstable and tends to release energy that can be turned into work.

\section{Heat flow in and out of $\sys$: Clausius inequality \label{sec:clausius}}
$\Delta \mathcal{Q}^{R}$ contains the total heat exchanged between $\sys$ and $\env$ due to resonant processes but, for our purposes, it is important to distinguish between the heat flowing from $\env$ to $\sys$ and the one flowing from $\sys$ to $\env$, which we respectively denote as $\Delta \mathcal{Q}_{\env \rightarrow \sys}^{R}$ and $\Delta \mathcal{Q}_{\env \leftarrow \sys}^{R}$, and satisfy $\Delta \mathcal{Q}^{R} = \Delta \mathcal{Q}_{\env \rightarrow \sys}^{R} + \Delta \mathcal{Q}_{\env \leftarrow \sys}^{R}$. To find $\Delta \mathcal{Q}_{\env \rightleftarrows \sys}^{R}$ we should notice that, as the reservoirs do not interact directly but rather through $\sys$, for an excitation to be transported from $\env_{\alpha}$ to $\env_{\beta}$ first it must go through the working medium. For example, in the process in which an excitation from a mode of frequency $\omega + k \omega_{d}$ in $\env_{\alpha}$ is transported to a mode of frequency $\omega$ in $\env_{\beta}$, a packet of energy $\hbar (\omega + k \omega_{d})$ flows from $\env_{\alpha}$ to $\sys$ and then $\hbar \omega$ flows from $\sys$ to $\env_{\beta}$, while emitting $k \hbar \omega_{d}$ into the driving field. The reverse process, in which a packet of energy $k \hbar \omega_{d}$ is absorbed from the driving field, can be described analogously. Therefore, to identify $\Delta \mathcal{Q}_{\env \rightleftarrows \sys}^{R}$ we can add and subtract $\omega$ to the integrand in Eq. \eqref{qR} to write it as
\begin{equation}
    \begin{aligned}
        & \Delta \mathcal{Q}^{R} / \tau_{d} = \\
        & - \sum_{k, \alpha , \beta} \int_{0}^{\infty} d \omega \, \hbar \omega  p_{\alpha \beta}^{(k)} (\omega) [n_{\beta} (\omega) - n_{\alpha} (\omega + k \omega_{d})] \\
        & + \sum_{k, \alpha , \beta} \int_{0}^{\infty} d \omega \hbar ( \omega + k \omega_{d}) p_{\alpha \beta}^{(k)} (\omega) [n_{\beta} (\omega) - n_{\alpha} (\omega + k \omega_{d})].
    \end{aligned}
    \label{qRSplit}
\end{equation}
If the integrand in Eq. \eqref{qR} is negative, which happens when $n_{\alpha} (\omega + k \omega_{d}) > n_{\beta} (\omega)$, then a packet of energy $k \hbar \omega_{d}$ is being emitted into the driving field (a condition that is satisfied if and only if $\omega$ belongs to the interval $\mathcal{I}^{-} = \{ \omega \in \mathbb{R}_{>0} |  n_{\alpha} (\omega + k \omega_{d}) > n_{\beta} (\omega) \}$). Thus, in this case, the first term in the right-hand side of Eq. \eqref{qRSplit} represents the heat flowing from $\sys$ to $\env$ and the second one, the heat flowing from $\env$ to $\sys$. Conversely, if the integrand in Eq. \eqref{qR} is positive, then a packet of energy $k \hbar \omega_{d}$ will be absorbed from the driving field (this happens if and only if $\omega$ belongs to $\mathcal{I}^{+}$, which is the complement of $\mathcal{I}^{-}$). In this case, the first term in the right-hand side of Eq. \eqref{qRSplit} represents the heat flowing from $\env$ to $\sys$ and the second one, the heat flowing from $\sys$ to $\env$. Hence, we find that $\Delta \mathcal{Q}_{\env \rightleftarrows \sys}^{R}$ is given by
\begin{equation}
    \begin{aligned}
        & \Delta \mathcal{Q}_{\env \rightleftarrows \sys}^{R} / \tau_{d} = \\
        & - \sum_{k, \alpha , \beta} \int_{\mathcal{I}^{\pm}} d \omega \, \hbar \omega  p_{\alpha \beta}^{(k)} (\omega) [n_{\beta} (\omega) - n_{\alpha} (\omega + k \omega_{d})] \\
        & + ´\sum_{k, \alpha , \beta} \int_{\mathcal{I}^{\mp}} d \omega \hbar ( \omega + k \omega_{d}) p_{\alpha \beta}^{(k)} (\omega) [n_{\beta} (\omega) - n_{\alpha} (\omega + k \omega_{d})].
    \end{aligned}
    \label{J}
\end{equation}
The above expressions can be used to obtain a lower bound for the ratio between the energy flowing out and into $\sys$: $\Delta \mathcal{Q}_{\env \leftarrow \sys}^{R} /  \lvert \Delta \mathcal{Q}_{\env \rightarrow \sys}^{R} \rvert$. For this we will use the fact that, without loss of generality, $n_{\alpha}$ can be written as a strictly decreasing function of the dimensionless variable $\omega / \Omega_{\alpha} (\omega)$, where $\Omega_{\alpha}$ is an appropriately chosen positive function (for example, for thermal reservoirs $\Omega_{\alpha}$ is a constant: $\Omega_{\alpha} (\omega) = k_{b} T_{\alpha} / \hbar$). Using this we can bound the ratio by noticing that if $\omega \in \mathcal{I}^{-}$ then $\omega > [\Omega_{\beta} (\omega) / \Omega_{\alpha} (\omega + k \omega_{d})] (\omega + k \omega_{d})$, and that the opposite inequality holds when $\omega \in \mathcal{I}^{+}$:
\begin{equation}
	\Delta \mathcal{Q}_{\env \leftarrow \sys}^{R} /  \lvert \Delta \mathcal{Q}_{\env \rightarrow \sys}^{R} \rvert > \text{min} \{ m , 1/M \},
	\label{clausius}
\end{equation}
where $m = \text{min} \{ \Omega_{\alpha} (\omega) / \Omega_{\beta} (\omega + k \omega_{d}) \}$ and $M = \text{max} \{ \Omega_{\alpha} (\omega) / \Omega_{\beta} (\omega + k \omega_{d}) \}$ (see the derivation in Appendix \ref{sec:clausiusAp}). Equation \eqref{clausius} is a form of Clausius inequality that can be applied to nonthermal reservoirs and shows that the amount of energy that it is lost, $\Delta \mathcal{Q}_{\env \leftarrow \sys}^{R}$, cannot be arbitrarily small. We can cast this inequality in more familiar terms in the case of thermal reservoirs: $\Delta \mathcal{Q}_{\env \rightarrow \sys}^{R} / T_{\text{h}} + \Delta \mathcal{Q}_{\env \leftarrow \sys}^{R} / T_{\text{c}} > 0$, where $T_{\text{c}}$ and $T_{\text{h}}$ are the temperatures of the coldest and hottest reservoirs, respectively. This is nothing but the usual Clausius theorem applied to a nonequilibrium Carnot engine working between two reservoirs with temperatures $T_{\text{c}}$ and $T_{\text{h}}$.

\section{Efficiency \label{sec:efficiency}}
The efficiency $\eta$ of the machine is defined as the ratio between the work extracted from $\sys$ and the energy injected by $\env$ into $\sys$. The numerator in $\eta$ (i.e., the extracted work) is $\lvert \bar{W}_{\sys} \rvert = \lvert \Delta \mathcal{Q}^{NR} + \Delta \mathcal{Q}_{\env \leftarrow \sys}^{R} - \lvert \Delta \mathcal{Q}_{\env \rightarrow \sys}^{R} \rvert \rvert$ (notice that work is extracted when if $\bar{W}_{\sys} < 0$, which is satisfied if and only if the energy flowing from $\env$ to $\sys$ is greater than the one that flows from $\sys$ to $\env$). On the other hand, the denominator in $\eta$ (i.e., the energy injected by $\env$ into $\sys$) is $\lvert \Delta \mathcal{Q}_{\env \to \sys}^{R} \rvert$. Thus, we can write the following exact expression for $\eta$:
\begin{equation}
    \eta = \frac{\lvert \Delta \mathcal{Q}_{\env \rightarrow \sys}^{R} \rvert - \Delta \mathcal{Q}_{\env \leftarrow \sys}^{R} - \Delta \mathcal{Q}^{NR}}{\lvert \Delta \mathcal{Q}_{\env \rightarrow \sys}^{R} \rvert}.
    \label{eff}
\end{equation}
It is immediate that $\eta$ can be bounded by dropping the negative contribution of the nonresonant processes from the numerator. Doing this, and using Eq. \eqref{clausius} we find that
\begin{equation}
    \eta < 1 - \frac{\Delta \mathcal{Q}_{\env \leftarrow \sys}^{R}}{\lvert \Delta \mathcal{Q}_{\env \rightarrow \sys}^{R} \rvert} < 1 - \text{min} \{ m , 1/M \},
    \label{effCarnot}
\end{equation}
which is a generalized bound for the efficiency of quantum linear engines. If the reservoirs are thermal $m = 1/M = T_{c}/T_{h}$, and we recover the Carnot limit: $\eta < 1 - T_{\text{c}} / T_{\text{h}}$. For other type of reservoirs, this bound can be violated (see below). The fact that the Carnot efficiency is recovered when the initial state of the reservoirs is a product of thermal states is a general result that goes beyond the limits of this model, and can be obtained, for example, by using fluctuation relations \cite{campisi, campisiPekolaFazio}. It is worth noticing that the nonresonant processes, which prevent the engine from reaching efficiencies arbitrarily close to the bound in Eq. \eqref{effCarnot}, are the same ones enforcing the third law of thermodynamics in the form of Nernst's unattainability principle \cite{freitasPaz17,freitasPaz18}.

\section{The cost of preparing nonthermal reservoirs \label{sec:cost}}
\begin{figure}[htp]
    \centering
    \includegraphics[scale=.65]{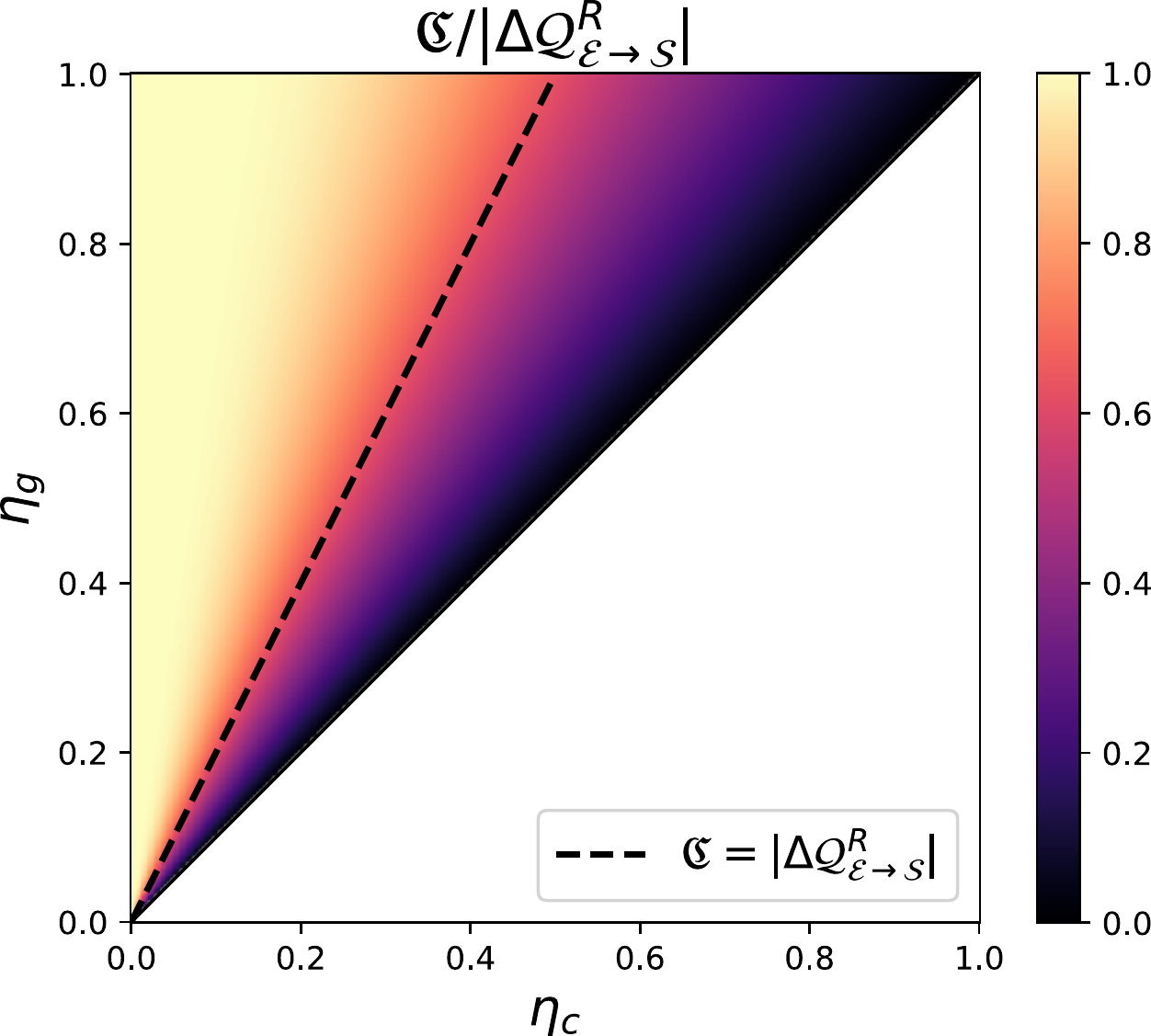}
    \caption{Normalized plot of $\mathfrak{C} / \lvert \Delta \mathcal{Q}_{\env \rightarrow \sys}^{R} \rvert$ as a function of $\eta_{c}$ and the target efficiency $\eta_{g}$, showing the different operating regimes. Anything in between the lower bound ($\eta_{g} = \eta_{c}$) and the dashed line ($\eta_{g} = 2 \eta_{c}$) belongs to the low-cost regime where improving the efficiency is cheaper than running the engine ($\mathfrak{C} < \lvert \Delta \mathcal{Q}_{\env \rightarrow \sys}^{R} \rvert$), while any improvement above the dashed line requires more energy to achieve ($\mathfrak{C} > \lvert \Delta \mathcal{Q}_{\env \rightarrow \sys}^{R} \rvert$).}
    \label{fig:cost}
\end{figure}

As shown above, linear quantum engines coupled to thermal reservoirs always satisfy the Carnot bound as their efficiency $\eta$ is less than $\eta_{c} = 1 - T_{\text{c}} / T_{\text{h}}$. Considering this holds independently of the evolution of the machine, one could argue that it should also be valid for a situation where: (i) thermal reservoirs are initially prepared, (ii) these reservoirs become nonthermal by driving them with an appropriately chosen quadratic Hamiltonian, and (iii) these nonthermal reservoirs are coupled to the working medium and used to run the engine. If the efficiency is computed only for stage (iii) of the above process one could observe an apparent violation of the Carnot bound as $\eta < \eta_{g} = 1 - \text{min} \{ m , 1/M \}$. Clearly, the difference between $\eta_{g}$ and $\eta_{c}$ comes from the energy invested in stage (ii), which is necessary to prepare the nonthermal reservoirs. Using our previous results, we can obtain this energetic cost $\mathfrak{C}$ relative to the energy required to run the engine as
\begin{equation}
	\mathfrak{C} / \lvert \Delta \mathcal{Q}_{\env \rightarrow \sys}^{R} \rvert \simeq (\eta_{g} - \eta_{c}) / \eta_{c}
	\label{cost}
\end{equation}
(see the derivation in Appendix \ref{sec:costAp}). Equation \eqref{cost} has two important consequences. First, it allows us to know \textit{a priori} the minimum cost to pay for a target work output and efficiency given some initial thermal resources. Indeed, using that $\lvert \bar{W}_{\sys} \rvert / \lvert \Delta \mathcal{Q}_{\env \rightarrow \sys}^{R} \rvert < \eta_{g}$ we obtain $\mathfrak{C} > \lvert \bar{W}_{\sys} \rvert (\eta_{g} - \eta_{c}) / \eta_{g} \eta_{c}$. Second, the energetic cost to improve the efficiency through the use of nonthermal resources relative to the energy needed to run the engine has a maximum possible value determined only by the temperatures of the coldest and hottest reservoirs: since $\eta_{g } \leq 1$, we find that $\mathfrak{C} / \lvert \Delta \mathcal{Q}_{\env \rightarrow \sys}^{R} \rvert \leq T_{c}/(T_{h} - T_{c})$. This means that, as $\eta_{g}$ approaches unity, $\mathfrak{C}$ grows as fast as $\lvert \Delta \mathcal{Q}_{\env \rightarrow \sys}^{R} \rvert$. Thus, improving the efficiency by using nonthermal resources is not much more expensive than running the engine. In fact, if we start with thermal reservoirs such that $T_{h} > 2 T_{c}$, as this implies that $\mathfrak{C}$ is smaller than $\lvert \Delta \mathcal{Q}_{\env \rightarrow \sys}^{R} \rvert$, we enter a low-cost regime where improving the efficiency is actually cheaper than running the engine (see Fig. \ref{fig:cost} for a phase diagram of the different operating regimes). For example, let us consider an engine working between a squeezed thermal reservoir at temperature $T_{h}$ and squeezing $r$ and a colder thermal reservoir at temperature $T_{c}$. We find that its efficiency is bounded by $\eta_{g} = 1 - T_{c} / \text{cosh} (2 r) T_{h}$ (in the limit of high $T_{h}$), and the cost to squeeze the hot reservoir is $\mathfrak{C} / \lvert \Delta \mathcal{Q}_{\env \rightarrow \sys}^{R} \rvert \simeq T_{c} [1 - 1/\text{cosh}(2r)] / (T_{h} - T_c)$ ($\Omega_{\alpha}$ for squeezed thermal reservoirs is computed as $\text{coth} ( \omega / \Omega_{\alpha}) = \text{cosh} (2 r_{\alpha}) \text{coth} (\hbar \omega / 2 k_{b} T_{\alpha})$). For this cost to be less than the energy needed to run the energy, the squeezing must be such that $2 - T_{h}/T_{c} < 1/\text{cosh(2r)}$. Thus, if $T_{h} < 2 T_{c}$, only small values of $r$ fulfill this inequality and high efficiencies cannot be reached in the low cost regime. Conversely, if $T_{h} > 2 T_{c}$, then $\eta_{g}$ can be arbitrarily close to one and $\mathfrak{C}$ would still be smaller than $\lvert \Delta \mathcal{Q}_{\env \rightarrow \sys}^{R} \rvert$. Finally, it is interesting to notice that, contrary to the case of large squeezing in which $\mathfrak{C}$ and $\lvert \Delta \mathcal{Q}_{\env \rightarrow \sys}^{R} \rvert$ scale similarly, for small values of $r$, as $\mathfrak{C} \simeq 2 T_{c} r^{2} \lvert \Delta \mathcal{Q}_{\env \rightarrow \sys}^{R} \rvert_{r=0} / (T_{h} - T_{c})$, $\mathfrak{C}$ grows as $r^{2}$ while $\lvert \Delta \mathcal{Q}_{\env \rightarrow \sys}^{R} \rvert$ remains approximately constant.

\section{The model \label{sec:model}}
Here we present a brief overview of the model, which is described in detail in Appendix \ref{sec:formalism}. As mentioned above, the dynamics of $\sys$ is governed by $M$ and $V(t)$, and each $\env_{\alpha}$ is a collection of noninteracting harmonic oscillators whose masses and frequencies are $m_{\alpha}$ and $\omega_{\alpha}$, respectively. The interaction Hamiltonian is $H_{\sys, \env} = \sum_{\alpha }X^{T} C_{\alpha} q_{\alpha}$, where $C_{\alpha}$ contains the coupling constants between $\sys$ and $\env_{\alpha}$. In Ref. \cite{freitasPaz17} it was shown that the solution of the Heisenberg equations for $X$ and $q_{\alpha}$ can be expressed as linear combinations of the corresponding Schr\"odinger operators, where the coefficients are determined by the dressed Green's function $G (t ,t^{\prime})$ of $\sys$, which is the retarded solution of
\begin{equation}
    M \partial_{t}^{2} G + V_{R} (t) G + \int_{0}^{t} d\tau \gamma (t - \tau) \partial_{\tau} G (\tau , t^{\prime}) = \delta (t-t^{\prime}).
    \label{eqG}
\end{equation}
Above, $\gamma (t) = \sum_{\alpha} \int d \omega I_{\alpha} ( \omega) \, \text{cos} (\omega t) / \omega$ is the dissipation kernel with $I_{\alpha} (\omega) = C_{\alpha} \delta (\omega - \omega_{\alpha}) C_{\alpha}^{T} / m \omega$ the spectral density of each $\env_{\alpha}$, and $V_{R}(t) = V(t) - \gamma(0)$ is the renormalized potential. For periodic $V$ we use Floquet theory to write $G (t , t^{\prime}) = \textstyle{\sum_{n = - \infty}^{+ \infty}} A_{n} ( t - t^{\prime})  e^{i n \omega_{d} t}$, and use Eq. \eqref{eqG} to obtain a set of linear differential equations for $A_{n} (t)$ and a corresponding set of linear algebraic equations for their Laplace transform $\tilde{A}_{n} (s)$ (see Appendix \ref{sec:formalism}). From the solution of the Heisenberg equations we can compute the long-time average of $d \langle H_{\env_{\alpha}}\rangle / dt$ and show that it only depends on the spectral densities and the coefficients $\tilde{A}_{n}$. $\Delta \mathcal{Q}$ is obtained by adding the contributions from all $\env_{\alpha}$'s. After a few manipulations (see Appendix \ref{sec:energy}), the expressions for $\Delta \mathcal{Q}^{NR}$ and $\Delta \mathcal{Q}^{R}$ shown in Eqs. \eqref{qNR} and \eqref{qR} respectively are obtained with $\tilde{p}_{\alpha \beta}^{(k)} (\omega) = (\pi/2) \text{tr}[I_{\alpha}(k \omega_{d} - \omega) \tilde{A}_{-k} (i \omega) I_{\beta}(\omega) \tilde{A}_{-k}^{\dagger} (i \omega)]$ and $p_{\alpha \beta}^{(k)} (\omega) = (\pi/2) \text{tr}[I_{\alpha}(\omega + k \omega_{d}) \tilde{A}_{k} (i \omega) I_{\beta}(\omega) \tilde{A}_{k}^{\dagger} (i \omega)]$. The existence of a high frequency cutoff in the spectral densities implies that both rates decrease with increasing $k$, guaranteeing the convergence of the series shown above.

\section{Conclusions \label{sec:conclusions}}
We presented the general theory of linear quantum engines, proving that these machines satisfy a form of Clausius inequality that prevents the amount of energy that it is lost and not converted into work from being arbitrarily small. As a consequence, their efficiency satisfy a generalized bound, which coincides with the well-known Carnot limit when coupled to thermal reservoirs and can be greater when coupled to nonthermal ones. We were able to estimate the energetic cost of transforming thermal reservoirs into nonthermal ones, which allows the engine to reach higher efficiencies by exploiting those nonthermal resources. We showed that this cost can be smaller than the energy needed to run the engine, even if the efficiency is arbitrarily close to one, and provided a lower bound to it in terms of a desired work output and efficiency. 

This work was supported by CONICET, UBACyT, and Agencia I+D+i.

\onecolumngrid
\appendix

\section{Formalism \label{sec:formalism}}

\subsection{Equations of motion and their solution}

In this section we will arrive at explicit formulas for the time evolution of the Heisenberg operators $X(t)$ and $q_{\alpha} (t)$ (we will use vector notation for the collection of position and momentum operators). From now on we will use units such that $k_{b} = \hbar = 1$. Our model includes a system $\sys$ consisting of a network of coupled driven oscillators, which is coupled to an environment $\env$ composed of different pieces $\env_{\alpha}$. The total Hamiltonian is $H = H_{\mathcal{S}} + H_{\mathcal{E}} + H_{\mathcal{S, E}}$ where
\begin{equation}
    \begin{aligned}
        & H_{\mathcal{S}} = P^{T} M^{-1} P / 2 + X^{T} V(t) V / 2 \\
        & H_{\mathcal{E}} = \sum_{\alpha} p_{\alpha}^{T} m_{\alpha}^{-1} p_{\alpha} / 2 + q_{\alpha}^{T} m_{\alpha} \omega_{\alpha}^{2} q_{\alpha} / 2 \\
        & H_{\mathcal{S, E}} = \sum_{\alpha} X^{T} C_{\alpha} q_{\alpha}
	\end{aligned}
	\label{hamiltonians}
\end{equation}
Above, $M$, $V(t)$, $m_{\alpha}$, and $\omega_{\alpha}$ are real symmetric matrices, with $m_{\alpha}$ and $\omega_{\alpha}$ diagonal matrices and $V(t)$ is $\tau_{d}$ periodic with driving frequency $\omega_{d} = 2 \pi / \tau_{d}$. Also, $C_{\alpha}$ is a real matrix containing the coupling constants between $\sys$ and $\env_{\alpha}$. The Heisenberg equations of motion for the position and momentum operators can be written as a system of second-order differential equations for the position operators only:
\begin{equation}
    \begin{aligned}
        \ddot{q}_{\alpha} + \omega_{\alpha}^{2} q_{\alpha} & = - m_{\alpha}^{-1} C_{\alpha}^{T} X \\
        M \ddot{X} + V(t) X & = - \sum_{\alpha} C_{\alpha} X
    \end{aligned}
    \label{position}
\end{equation}
The solution of the first equation in Eq. \eqref{position}, which corresponds to the position operators of $\env_{\alpha}$, is
\begin{equation}
    q_{\alpha} (t) = q_{\alpha}^{h} (t) - (m_{\alpha} \omega_{\alpha})^{-1} \int_{0}^{t} dt^{\prime} \, \text{sin} [\omega_{\alpha} (t - t^{\prime})] C_{\alpha}^{T} X (t^{\prime})
    \label{solutionQ}
\end{equation}
where $q_{\alpha}^{h}(t) = \text{cos} ( \omega_{\alpha} t ) q_{\alpha,0} + ( m_{\alpha} \omega_{\alpha} )^{-1} \text{sin} ( \omega_{\alpha} t ) p_{\alpha,0}$ is the solution of the homogeneous equation with $q_{\alpha,0}$ and $p_{\alpha,0}$ being the corresponding Schr\"odinger operators. In order to obtain the solution for the position operators of $\sys$, $X$, we replace Eq. \eqref{solutionQ} in the second equation in Eq. \eqref{position}:
\begin{equation}
	M \ddot{X} + V (t) X - \int_{0}^{t} d \tau \left[ \sum_{\alpha} C_{\alpha} (m_{\alpha} \omega_{\alpha})^{-1} \text{sin} [\omega_{\alpha} (t - \tau)] C_{\alpha}^{T} \right] X ( \tau ) = - \sum_{\alpha} C_{\alpha} q_{\alpha}^{h}.
	\label{eqX}
\end{equation}
Equation \eqref{eqX} is a linear second order inhomogeneous integro-differential equation and, thus, we will solve it using the Green's function method. In particular, we consider its retarded Green's function $G$, which is the unique solution of
\begin{equation}
	M \frac{\partial^{2} G}{\partial t^{2}} (t,t^{\prime}) + V (t) G (t,t^{\prime}) - \int_{0}^{t} d \tau \left[ \sum_{\alpha} C_{\alpha} (m_{\alpha} \omega_{\alpha})^{-1} \text{sin} [\omega_{\alpha} (t - \tau)] C_{\alpha}^{T} \right] G ( \tau , t^{\prime} ) = \delta (t - t^{\prime})
	\label{eqXGreen}
\end{equation}
with initial conditions $G (t^{\prime} , t^{\prime}) = 0$ and $\partial_{t} G (t^{\prime} , t^{\prime}) = M^{-1}$. After performing a partial integration in Eq. \eqref{eqXGreen} we arrive at Eq. \eqref{eqG} shown in the main text
\begin{equation}
	M \frac{\partial^{2} G}{\partial t^{2}} (t,t^{\prime}) + V_{R} (t) G (t,t^{\prime}) - \int_{0}^{t} d \tau \, \gamma (t - \tau) \, \frac{\partial G}{\partial \tau} ( \tau , t^{\prime} ) = \delta (t - t^{\prime}),
	\label{eqXGreen2}
\end{equation}
where we defined the dissipation kernel $\gamma (t) = \sum_{\alpha} \int d \omega I_{\alpha} ( \omega) \, \text{cos} (\omega t) / \omega$ with $I_{\alpha} (\omega) = C_{\alpha} \delta (\omega - \omega_{\alpha}) C_{\alpha}^{T} / m \omega$ the spectral density of each $\env_{\alpha}$, and $V_{R}(t) = V(t) - \gamma(0)$ is the renormalized potential. Finally, the solution for $X$ is
\begin{equation}
    X (t) = X^{h} (t) - \int_{0}^{t} d t^{\prime} \, G(t , t^{\prime}) \sum_{\alpha} C_{\alpha} q_{\alpha}^{h} (t^{\prime}),
    \label{solutionX}
\end{equation}
with $X^{h} = F(t,0)X_{0} + G(t,0) P_{0}$ the homogeneous solution of Eq. \eqref{eqX} ($F$ is the advanced Green function of Eq. \eqref{eqX}, which is computed in a similar way, and $X_{0}$ and $P_{0}$ are the corresponding Schr\"odinger operators). With Eqs. \eqref{solutionQ} and \eqref{solutionX} we can write the time evolution of both $q_{\alpha}$ and $X$ as linear combinations of Schr\"odinger operators with time-dependent coefficients, which are functions of $G$ only. This proves the statement that was included in the description of the model appearing in the main text.

\subsection{Solution for the case of a periodic potential}

For periodic $V(t)$ such that $V_{R} (t) = \sum_{n} V_{n} e^{i n \omega_{d} t}$ we can write $G(t,t^{\prime}) = \sum_{n=-\infty}^{n=+\infty} A_{n} (t -t^{\prime}) e^{i n \omega_{d} t}$, were $A_{n} (t<0) = 0$. Replacing this in Eq. \eqref{eqXGreen2} we obtain a set of coupled linear integro-differential equations for the coefficients $A_{n} (t)$. This set of equations can be conveniently expressed as a linear set of coupled algebraic equations for their Laplace transform $\tilde{A}_{n} (s)$, which reads:
\begin{equation}
	\tilde{g}^{-1} \left( s + i n \omega_{d} \right) \, \tilde{A}_{n} \left( s \right) + \sum_{m \neq 0} V_{m} \tilde{A}_{n-m} \left( s \right) = \delta_{n0} \mathbb{1}.
	\label{eqLaplace}
\end{equation}
Above, $\tilde{g}^{-1} \left( s \right) = M s^{2} + \omega_{r}^{2} + s \, \tilde{\gamma} \left( s \right)$ is the static Green's function with renormalized frequency $\omega_{r}^{2} = V_{0} - \gamma(0)$. Equation \eqref{eqLaplace} is particularly useful because, as we will see below, in the long-time limit all the relevant quantities can be written in terms of the coefficients $\tilde{A}_{n}$ (instead of $A_{n}$). Thus, it is not necessary to use the inverse Laplace transform to compute $A_{n} (t)$.

\subsection{Properties of the Green's function}

Here we will discuss three important properties of the Green's function $G$ in terms of the Laplace coefficients $\tilde{A}_{n}$. These are: the complex conjugation of $\tilde{A}_{n}$, a property that relates $\tilde{A}_{n}$ evaluated in two different arguments, and the long-time behavior of $A_{n}$. The first and second properties will prove useful below when we derive the heat currents shown in the main text, while the third one is related to the existence of a stable stationary regime for $\sys$.

\begin{itemize}
    \item \textbf{Complex conjugation}: Since $G$ is a real valued function, it follows that $\tilde{A}_{n} (s) = \tilde{A}_{-n}^{\ast} (s^{\ast})$.
    \item \textbf{Translation of the argument}: Here we will prove the validity of
    \begin{equation}
        \tilde{A}_{n} (s) = \tilde{A}_{-n}^{T} (s + in \omega_{d}),
    \end{equation}
    for any driving satisfying $V(t) = V(-t)$. In what follows we will use two vectors and a matrix. The first vector is $\tilde{A} (s)$, whose $n$-th component coincides with $\tilde{A}_{n}$. The second one is $\mathcal{I}_{n}$, whose components are all zero except for the $n$-th one, which coincides with the identity matrix. In other words, $\mathcal{I}_{n}$ is the $n$-th element of the canonical basis of the vector space where $\tilde{A}(s)$ lives. The matrix is $\mathcal{M}_{m,n} (s) = \tilde{g}^{-1} (s + i m \omega_{d}) \delta_{m,n} + (1 - \delta_{m,n}) V_{m-n}$. Then, Eq. \eqref{eqLaplace} can be written as $\mathcal{M} (s) \tilde{A} (s) = \mathcal{I}_{0}$. Assuming $\mathcal{M}$ is invertible, the solution of the previous equation is
    \begin{equation}
        \tilde{A}_{n} (s) = \mathcal{I}_{n}^{T} \mathcal{M}^{-1} (s) \mathcal{I}_{0}.
        \label{tras1}
    \end{equation}
    The first step to arrive at the desired result is writing Eq. \eqref{eqLaplace} but with the argument translated an amount $il \omega_{d}$, i.e., $s \rightarrow s + i l \omega_{d}$:
    \begin{equation}
        \sum_{n} \mathcal{M}_{m,n} (s + il \omega_{d}) \tilde{A}_{n} (s + il \omega_{d}) = \delta_{m,0} \mathbb{1}
        \label{tras2}
    \end{equation}
    Using the fact that $\mathcal{M}_{m,n} (s + il \omega_{d}) = \mathcal{M}_{m+l,n+l} (s)$ and defining new indices $n^{\prime} = n + l$ and $m^{\prime} = m + l$, the previous equation reads
    \begin{equation}
        \sum_{n^{\prime}} \mathcal{M}_{m^{\prime},n^{\prime}} (s) \tilde{A}_{n^{\prime} - l} (s + il \omega_{d}) = \delta_{m^{\prime},l} \mathbb{1}.
        \label{tras3}
    \end{equation}
    Without loss of generality, we can assume that $V(t) = V(-t)$. This is because we are solving an initial value problem (where initial values are given at $t=0$) and can extend the function $V(t)$ to negative values in an arbitrary way. If this is the case, then $V_{n}^{\ast} = V_{-n} = V_{n}$ and, thus, $\mathcal{M}_{m,n} (s) = \mathcal{M}_{n,m} (s)$. Therefore, Eq. \eqref{tras3} can be written as
    \begin{equation}
        \sum_{n} \mathcal{M}_{n,m} (s) \tilde{A}_{n - l} (s + il \omega_{d}) = \delta_{m-l,0} \mathbb{1},
        \label{tras4}
    \end{equation}
    where we dropped the tilde in the indices. In order to write Eq. \eqref{tras4} in matrix form, we define the column vector $\tilde{A}_{\updownarrow} (s)$ as the one that contains $\tilde{A}_{n - l} (s)$ in the $n$-th row. That is, it is displaced $l$ rows compared to the vector $\tilde{A} (s)$. Then, Eq. \eqref{tras4} reads
    \begin{equation}
        \mathcal{M}^{T} (s) \tilde{A}_{\updownarrow} (s + il \omega_{d}) = \mathcal{I}_{l}.
        \label{tras5}
    \end{equation}
    Its solution is obtained by inverting $\mathcal{M}^{T}$ and multiplying by $\mathcal{I}_{0}^{T}$ from the left:
    \begin{equation}
        \tilde{A}_{-n} (s + in \omega_{d}) = \mathcal{I}_{0}^{T} \mathcal{M}^{T - 1} (s) \mathcal{I}_{n}
        \label{tras6}
    \end{equation}
    Comparing Eqs. \eqref{tras1} and \eqref{tras6}, we conclude that
    \begin{equation}
        \tilde{A}_{n} (s) = \tilde{A}_{-n}^{T} (s + in \omega_{d}).
    \end{equation}
    \item \textbf{Long-time behavior}: Assuming the limit $\lim_{t \to \infty} A_{n} \left( t \right) \equiv A_{n} \left( \infty \right)$ exists and it is finite (which is not the case for environments with a discrete number of modes), then due to the Laplace final value theorem we have $A_{n} \left( \infty \right) = \lim_{s \to 0} s \, \tilde{A}_{n} \left( s \right)$. Plugging this in Eq. \eqref{eqLaplace} we arrive at
    \begin{equation}
        \sum_{k} \mathcal{M}_{m,n} (0) A_{n} (\infty) = 0,
    \end{equation}
    where $\mathcal{M}$ is the matrix defined above. If $\mathcal{M} (0)$ is invertible, then the only solution is $A_{n} (\infty) = 0$, meaning that the Green function $G$ decays for long times. One possible criteria to decide about the invertibility of $\mathcal{M} (0)$ is to check if it is a strictly block diagonally dominant matrix (generalized Levy-Desplanques theorem). This condition can be written as
    \begin{equation}
       \text{min}_{n} \{ \| \tilde{g}(i n \omega_{d}) \|^{-1} \} > \sum_{m \neq 0} \| V_{m} \|,
    \end{equation}
    where $\| \cdot \|$ is the usual operator norm. The above inequality compares the effect of the dissipation induced by the environment (the dissipation kernel dominates the left-hand side) and the effect of the driving (which determines the right-hand side). Therefore, the invertibility is satisfied if the dissipation is strong enough to enforce the inequality.
\end{itemize}

\subsection{Unitarity}

In this section we write a property that will be useful for later computations and it is the consequence of the unitarity of the temporal evolution. Unitarity implies that the commutations relations are preserved: $[q_{\alpha} (t) , p_{\alpha}^{T} (t)] = i \mathbb{1} \, \forall t$. Imposing this to the solutions of the Heisenberg equations presented above, we arrive at the following condition for the Green's function $G$ that must be satisfied at all times:
\begin{equation}
    \begin{aligned}
        0 & = \frac{1}{2} \, \text{Im} \left[ e^{i \omega_{\alpha} t} \int_{0}^{t} dt^{\prime} \, e^{- i \omega_{\alpha} t^{\prime}} C_{\alpha}^{T} \int_{0}^{t^{\prime}} dt^{\prime \prime} \, G (t^{\prime} , t^{\prime \prime}) C_{\alpha} e^{i \omega_{\alpha} t^{\prime \prime}} e^{-i \omega_{\alpha} t} \right] \\
        & + \frac{1}{2} \, \text{Im} \left[ e^{-i \omega_{\alpha} t} \int_{0}^{t} dt^{\prime} \int_{0}^{t^{\prime}} dt^{\prime \prime} \, e^{i \omega_{\alpha} t^{\prime \prime}} C_{\alpha}^{T}  \, G^{T} (t^{\prime} , t^{\prime \prime}) C_{\alpha} e^{- i \omega_{\alpha} t^{\prime}} e^{i \omega_{\alpha} t} \right] \\
        & - \sum_{\beta} \int_{0}^{t} dt^{\prime} \int_{0}^{t} dt^{\prime \prime} \text{sin} [ \omega_{\alpha} (t - t^{\prime}) ] \\
        & \times C_{\alpha}^{T} \text{Im} \left[ \int_{0}^{t^{\prime}} d t_{1} \int_{0}^{t^{\prime \prime}} d t_{2} \, G (t^{\prime},t_{1}) C_{\beta} e^{i \omega_{\beta} t_{1}} \left( m_{\beta} \omega_{\beta} \right)^{-1} \, e^{- i \omega_{\beta} t_{2}} C_{\beta}^{T} G^{T} (t^{\prime \prime},t_{2}) \right] C_{\alpha} \text{cos} [ \omega_{\alpha} (t - t^{\prime \prime}) ] \\
        & - \int_{0}^{t} dt^{\prime} \int_{0}^{t} dt^{\prime \prime} \, \text{sin} [ \omega_{\alpha} (t - t^{\prime}) ] C_{\alpha}^{T} \langle i [X^{h} (t^{\prime}) , X^{h T} (t^{\prime \prime})] \rangle C_{\alpha} \text{cos} [ \omega_{\alpha} (t - t^{\prime \prime}) ].
    \end{aligned}
    \label{unitary}
\end{equation}
As discussed in Ref. \cite{aguilarFreitasPaz}, in the long-time limit this equation can be interpreted as form of the fluctuation-dissipation relation.

\section{Energy and heat currents \label{sec:energy}}

\subsection{Mean energy of a reservoir}

Here we will compute the mean energy of one thermal reservoir, $\langle H_{\env_{\alpha}} \rangle$, using the solutions of the Heisenberg equations presented in Appendix \ref{sec:formalism}. In the sections below, we will analyze its long-time limit and then obtain the heat currents from it. The mean value of $H_{\env_{\alpha}}$ reads:
\begin{equation}
    \langle H_{\env_{\alpha}} \rangle = \text{tr} [m_{\alpha}^{-1} \langle \{ p_{\alpha} , p_{\alpha}^{T} \} \rangle] / 4 + \text{tr} [m_{\alpha} \omega_{\alpha}^{2} \langle \{ q_{\alpha} , q_{\alpha}^{T} \} \rangle] / 4.
\end{equation}
In order to compute this, we need to specify the initial state of the environments. We will consider arbitrary initial states such that
\begin{equation}
	\begin{aligned}
		\langle \{ q_{\alpha} , q_{\beta}^{T} \} \rangle & = (m_{\alpha} \omega_{\alpha})^{-1} A_{\alpha}(\omega_{\alpha}) \delta_{\alpha \beta} \\
		\langle \{ p_{\alpha} , p_{\beta}^{T} \} \rangle & = m_{\alpha} \omega_{\alpha} \, B_{\alpha}(\omega_{\alpha}) \delta_{\alpha \beta} \\
		\langle \{ q_{\alpha} , p_{\beta}^{T} \} \rangle & = D_{\alpha}(\omega_{\alpha}) \delta_{\alpha \beta},
	\end{aligned}
\end{equation}
where $A_{\alpha}$ and $B_{\alpha}$ are positive functions, and $D_{\alpha}$ is an arbitrary function. $A_{\alpha}$, $B_{\alpha}$ and $D_{\alpha}$ not only depend on the frequency of the mode, but also on various parameters of $\env_{\alpha}$ (such as the temperature and squeezing, for example). In order to simplify expressions we define two functions $f_{\alpha}$ and $g_{\alpha}$ as
\begin{equation}
	\begin{aligned}
		f_{\alpha} (\omega) & = [A_{\alpha}(\omega) + B_{\alpha}(\omega)] / 2 \\
		g_{\alpha} (\omega) & = [A_{\alpha}(\omega) - B_{\alpha}(\omega)] / 2.
	\end{aligned}
\end{equation}
Before continuing, note that
\begin{equation}
	\langle H_{\env_{\alpha}} (t=0) \rangle = \frac{1}{2} \text{tr} [ \omega_{\alpha} \, f_{\alpha} (\omega_{\alpha}) ]
\end{equation}
and, therefore, we can write $f_{\alpha} (\omega) = 2 n_{\alpha} (\omega) + 1$, where $n_{\alpha} (\omega)$ is the mean value of the number operator corresponding to the mode of frequency $\omega$ in $ \env_{\alpha}$ at $t=0$. Also, since $f_{\alpha}$ is defined for positive frequencies only, we can assume, without loss of generality, that $f_{\alpha} (- \omega) = -  f_{\alpha} (\omega)$. With the definitions above, we have
\begin{equation}
		\langle \{ q_{\alpha}^{h} (t_{1}) , q_{\beta}^{h T} (t_{2}) \}  \rangle = (m_{\alpha} \omega_{\alpha} )^{-1} \left\{ f_{\alpha} ( \omega_{\alpha} ) \text{Re} [ e^{i \omega_{\alpha} (t_{1} - t_{2})} ] + g_{\alpha} ( \omega_{\alpha}) \text{Re} [ e^{i \omega_{\alpha} (t_{1} + t_{2})} ] + D_{\alpha} ( \omega_{\alpha} ) \, \text{Im} [ e^{i \omega_{\alpha} (t_{1} + t_{2})} ]  \right\}.
\end{equation}
Using the previous equation and the fact that $p_{\alpha} = m_{\alpha} \dot{q}_{\alpha}$, a straightforward computation leads to
\begin{equation}
    \begin{aligned}
        \langle H_{\mathcal{E}_{\alpha}} \rangle & = \frac{1}{2} \, \text{tr} [ \omega_{\alpha} \,  f_{\alpha} (\omega_{\alpha}) ] \\
        & + \frac{1}{4} \int_{0}^{t} dt^{\prime} \int_{0}^{t} dt^{\prime \prime} \, \text{tr} \left[ m_{\alpha}^{-1} \, \text{cos} [ \omega_{\alpha} (t^{\prime} - t^{\prime \prime})] \, C_{\alpha}^{T} \langle \{ X^{h} ( t^{\prime}) , X^{h T} ( t^{\prime \prime} ) \} \rangle C_{\alpha} \right] \\
        & + \frac{1}{2} \, \text{tr} \left\{ m_{\alpha}^{-1} \,  f_{\alpha} (\omega_{\alpha}) \, \text{Im} \left[ \int_{0}^{t} dt^{\prime} \, e^{- i \omega_{\alpha} t^{\prime}} C_{\alpha}^{T} \int_{0}^{t^{\prime}} dt^{\prime \prime} \, G (t^{\prime},t^{\prime \prime}) \, C_{\alpha} \, e^{i \omega_{\alpha} t^{\prime \prime}} \right] \right\} \\
        & - \frac{1}{2} \, \text{tr} \left\{ m_{\alpha}^{-1} \,  g_{\alpha} (\omega_{\alpha}) \, \text{Im} \left[ \int_{0}^{t} dt^{\prime} \, e^{i \omega_{\alpha} t^{\prime}} C_{\alpha}^{T} \int_{0}^{t^{\prime}} dt^{\prime \prime} \, G (t^{\prime},t^{\prime \prime}) \, C_{\alpha} \, e^{i \omega_{\alpha} t^{\prime \prime}} \right] \right\} \\
        & + \frac{1}{2} \, \text{tr} \left\{ m_{\alpha}^{-1} \,  D_{\alpha} (\omega_{\alpha})\, \text{Im} \left[ \int_{0}^{t} dt^{\prime} \, e^{i \omega_{\alpha} t^{\prime}} C_{\alpha}^{T} \int_{0}^{t^{\prime}} dt^{\prime \prime} \, G (t^{\prime},t^{\prime \prime}) \, C_{\alpha} \, e^{i \omega_{\alpha} t^{\prime \prime}} \right] \right\} \\
        & + \frac{1}{4} \sum_{\beta} \int_{0}^{t} dt^{\prime} \int_{0}^{t} dt^{\prime \prime} \, \text{tr} \left\{ m_{\alpha}^{-1} \, \text{cos} [ \omega_{\alpha} (t^{\prime} - t^{\prime \prime}) ] \right. \\
        & \left. \times C_{\alpha}^{T} \text{Re} \left[ \int_{0}^{t^{\prime}} d t_{1} \int_{0}^{t^{\prime \prime}} d t_{2} \, G (t^{\prime},t_{1}) C_{\beta} e^{i \omega_{\beta} t_{1}} ( m_{\beta} \omega_{\beta})^{-1} \,  f_{\beta} (\omega_{\beta}) \, e^{- i \omega_{\beta} t_{2}} C_{\beta}^{T} G^{T} (t^{\prime \prime},t_{2}) \right] C_{\alpha} \right\} \\
        & + \frac{1}{4} \sum_{\beta} \int_{0}^{t} dt^{\prime} \int_{0}^{t} dt^{\prime \prime} \, \text{tr} \left\{ m_{\alpha}^{-1} \, \text{cos} [ \omega_{\alpha} (t^{\prime} - t^{\prime \prime}) ] \right. \\
        & \left. \times C_{\alpha}^{T} \text{Re} \left[ \int_{0}^{t^{\prime}} d t_{1} \int_{0}^{t^{\prime \prime}} d t_{2} \, G (t^{\prime},t_{1}) C_{\beta} e^{i \omega_{\beta} t_{1}} ( m_{\beta} \omega_{\beta})^{-1} \,  g_{\beta} (\omega_{\beta}) \, e^{i \omega_{\beta} t_{2}} C_{\beta}^{T} G^{T} (t^{\prime \prime},t_{2}) \right] C_{\alpha} \right\} \\
        & + \frac{1}{4} \sum_{\beta} \int_{0}^{t} dt^{\prime} \int_{0}^{t} dt^{\prime \prime} \, \text{tr} \left\{ m_{\alpha}^{-1} \, \text{cos} [ \omega_{\alpha} (t^{\prime} - t^{\prime \prime}) ] \right. \\
        & \left. \times C_{\alpha}^{T} \text{Im} \left[ \int_{0}^{t^{\prime}} d t_{1} \int_{0}^{t^{\prime \prime}} d t_{2} \, G (t^{\prime},t_{1}) C_{\beta} e^{i \omega_{\beta} t_{1}} ( m_{\beta} \omega_{\beta})^{-1} \,  D_{\beta} (\omega_{\beta}) \, e^{i \omega_{\beta} t_{2}} C_{\beta}^{T} G^{T} (t^{\prime \prime},t_{2}) \right] C_{\alpha} \right\}
    \end{aligned}
    \label{energy1}
\end{equation}
We can rewrite the above expression using the unitarity of the evolution (Eq. \eqref{unitary}) and a simple consequence of the properties of the trace functional:
\begin{equation}
    \begin{aligned}
        & \text{tr} \left\{ m_{\alpha}^{-1} \,  f_{\alpha} (\omega_{\alpha}) \, \text{Im} \left[ \int_{0}^{t} dt^{\prime} \, e^{- i \omega_{\alpha} t^{\prime}} C_{\alpha}^{T} \int_{0}^{t^{\prime}} dt^{\prime \prime} \, G (t^{\prime},t^{\prime \prime}) \, C_{\alpha} \, e^{i \omega_{\alpha} t^{\prime \prime}} \right] \right\} \\
        & = \text{tr} \left\{ m_{\alpha}^{-1} \,  f_{\alpha} (\omega_{\alpha}) \left\{ \frac{1}{2} \, \text{Im} \left[  e^{i \omega_{\alpha} t} \int_{0}^{t} dt^{\prime} \, e^{- i \omega_{\alpha} t^{\prime}} C_{\alpha}^{T} \int_{0}^{t^{\prime}} dt^{\prime \prime} \, G (t^{\prime},t^{\prime \prime}) \, C_{\alpha} \, e^{i \omega_{\alpha} t^{\prime \prime}} e^{- i \omega_{\alpha} t} \right] \right. \right. \\
        & \left. \left. + \frac{1}{2} \, \text{Im} \left[  e^{- i \omega_{\alpha} t} \int_{0}^{t} dt^{\prime} \int_{0}^{t^{\prime}} dt^{\prime \prime} \, e^{i \omega_{\alpha} t^{\prime \prime}} C_{\alpha}^{T}  \, G^{T} (t^{\prime} , t^{\prime \prime}) C_{\alpha} e^{- i \omega_{\alpha} t^{\prime}} e^{i \omega_{\alpha} t} \right] \right\} \right\}.
    \end{aligned}
\end{equation}
Doing  this, Eq. \eqref{energy1} transforms into
\begin{equation}
    \begin{aligned}
        \langle H_{\mathcal{E}_{\alpha}} \rangle & = \frac{1}{2} \, \text{tr} \left[ \omega_{\alpha} \,  f_{\alpha} (\omega_{\alpha}) \right] \\
        & + \frac{1}{4} \int_{0}^{t} dt^{\prime} \int_{0}^{t} dt^{\prime \prime} \, \text{tr} \left\{ m_{\alpha}^{-1} \, \text{cos} \left[ \omega_{\alpha} \left(t^{\prime} - t^{\prime \prime} \right) \right] \, C_{\alpha}^{T} \langle \{ X^{h} \left( t^{\prime} \right) , X^{h \ T} \left( t^{\prime \prime} \right) \} \rangle C_{\alpha} \right\} \\
        & - \frac{1}{4} \int_{0}^{t} dt^{\prime} \int_{0}^{t} dt^{\prime \prime} \,  \text{tr} \left\{ m_{\alpha}^{-1} \,  f_{\alpha} (\omega_{\alpha}) \, \text{sin} [ \omega_{\alpha} (t^{\prime} - t^{\prime \prime}) ] \, C_{\alpha}^{T} \langle i \, [X^{h} (t^{\prime}) , X^{h T} (t^{\prime \prime})] \rangle C_{\alpha} \right\} \\
        & - \frac{1}{2} \, \text{tr} \left\{ m_{\alpha}^{-1} \,  g_{\alpha} (\omega_{\alpha}) \, \text{Im} \left[ \int_{0}^{t} dt^{\prime} \, e^{i \omega_{\alpha} t^{\prime}} C_{\alpha}^{T} \int_{0}^{t^{\prime}} dt^{\prime \prime} \, G (t^{\prime},t^{\prime \prime}) \, C_{\alpha} \, e^{i \omega_{\alpha} t^{\prime \prime}} \right] \right\} \\
        & + \frac{1}{2} \, \text{tr} \left\{ m_{\alpha}^{-1} \,  D_{\alpha} (\omega_{\alpha})\, \text{Im} \left[ \int_{0}^{t} dt^{\prime} \, e^{i \omega_{\alpha} t^{\prime}} C_{\alpha}^{T} \int_{0}^{t^{\prime}} dt^{\prime \prime} \, G (t^{\prime},t^{\prime \prime}) \, C_{\alpha} \, e^{i \omega_{\alpha} t^{\prime \prime}} \right] \right\} \\
        & - \frac{1}{4} \sum_{\beta} \text{tr} \left\{ m_{\alpha}^{-1} \,  f_{\alpha} (\omega_{\alpha}) \, \int_{0}^{t} dt^{\prime} \int_{0}^{t} dt^{\prime \prime} \, \text{sin} [ \omega_{\alpha} (t^{\prime} - t^{\prime \prime}) ] \right. \\
        & \times \left. C_{\alpha}^{T} \text{Im} \left[ \int_{0}^{t^{\prime}} d t_{1} \int_{0}^{t^{\prime \prime}} d t_{2} \, G (t^{\prime},t_{1}) C_{\beta} e^{i \omega_{\beta} t_{1}} \left( m_{\beta} \omega_{\beta} \right)^{-1} \, e^{- i \omega_{\beta} t_{2}} C_{\beta}^{T} G^{T} (t^{\prime \prime},t_{2}) \right] C_{\alpha} \right\} \\
        & + \frac{1}{4} \sum_{\beta} \int_{0}^{t} dt^{\prime} \int_{0}^{t} dt^{\prime \prime} \, \text{tr} \left[ m_{\alpha}^{-1} \, \text{cos} \left[ \omega_{\alpha} \left(t^{\prime} - t^{\prime \prime} \right) \right] \right. \\
        & \left. \times C_{\alpha}^{T} \text{Re} \left[ \int_{0}^{t^{\prime}} d t_{1} \int_{0}^{t^{\prime \prime}} d t_{2} \, G (t^{\prime},t_{1}) C_{\beta} e^{i \omega_{\beta} t_{1}} \left( m_{\beta} \omega_{\beta} \right)^{-1} \,  f_{\beta} (\omega_{\beta}) \, e^{- i \omega_{\beta} t_{2}} C_{\beta}^{T} G^{T} (t^{\prime \prime},t_{2}) \right] C_{\alpha} \right] \\
        & + \frac{1}{4} \sum_{\beta} \int_{0}^{t} dt^{\prime} \int_{0}^{t} dt^{\prime \prime} \, \text{tr} \left\{ m_{\alpha}^{-1} \, \text{cos} [ \omega_{\alpha} (t^{\prime} - t^{\prime \prime}) ] \right. \\
        & \left. \times C_{\alpha}^{T} \text{Re} \left[ \int_{0}^{t^{\prime}} d t_{1} \int_{0}^{t^{\prime \prime}} d t_{2} \, G (t^{\prime},t_{1}) C_{\beta} e^{i \omega_{\beta} t_{1}} ( m_{\beta} \omega_{\beta})^{-1} \,  g_{\beta} (\omega_{\beta}) \, e^{i \omega_{\beta} t_{2}} C_{\beta}^{T} G^{T} (t^{\prime \prime},t_{2}) \right] C_{\alpha} \right\} \\
        & + \frac{1}{4} \sum_{\beta} \int_{0}^{t} dt^{\prime} \int_{0}^{t} dt^{\prime \prime} \, \text{tr} \left\{ m_{\alpha}^{-1} \, \text{cos} [ \omega_{\alpha} (t^{\prime} - t^{\prime \prime}) ] \right. \\
        & \left. \times C_{\alpha}^{T} \text{Im} \left[ \int_{0}^{t^{\prime}} d t_{1} \int_{0}^{t^{\prime \prime}} d t_{2} \, G (t^{\prime},t_{1}) C_{\beta} e^{i \omega_{\beta} t_{1}} ( m_{\beta} \omega_{\beta})^{-1} \,  D_{\beta} (\omega_{\beta}) \, e^{i \omega_{\beta} t_{2}} C_{\beta}^{T} G^{T} (t^{\prime \prime},t_{2}) \right] C_{\alpha} \right\}
    \end{aligned}
    \label{energy2}
\end{equation}
We notice that the last two terms in this expression can be conveniently expressed in terms of the spectral density as
\begin{equation}
    \begin{aligned}
       C_{\beta} e^{i \omega_{\beta} t_{1}} \left( m_{\beta} \omega_{\beta} \right)^{-1} \, f_{\beta} (\omega_{\beta}) \, e^{- i \omega_{\beta} t_{2}} C_{\beta}^{T} = \int_{0}^{\infty} d \omega \, I_{\beta} ( \omega ) e^{i \omega (t_{1} - t_{2})} \, f_{\beta} (\omega).
    \end{aligned}
\end{equation}
In what follows it will be convenient to use the notation
\begin{equation}
    \mathcal{K} (\omega , \omega^{\prime} , t) = \int_{0}^{t} dt^{\prime} e^{- i \omega^{\prime} t^{\prime}} \int_{0}^{t^{\prime}} d t^{\prime \prime} G (t^{\prime} , t^{\prime \prime}) e^{i \omega t^{\prime \prime}}.
    \label{jFunction}
\end{equation}
In fact, it will be useful to express $\mathcal{K}$ in terms of the $A_{n}$ coefficients of the Green's function. After some calculations one can show that
\begin{equation}
	\mathcal{K} \left( \omega , \omega^{\prime} , t \right) = \sum_{n} \left[ t \, \text{sinc} [ ( \omega - \omega^{\prime} + n \omega_{d}) t/2 ] \, a_{n} ( i \omega ) \, e^{i ( \omega - \omega^{\prime} + n \omega_{d} ) t/2} + F_{n} ( \omega , \omega^{\prime} ) \right],
\end{equation}
where we defined
\begin{equation}
	\begin{aligned}
	a_{n} ( i \omega ) & = \int_{0}^{t} d t^{\prime} \, A_{n} ( t^{\prime} ) \, e^{- i \omega t^{\prime}} \\
	F_{n} ( \omega , \omega^{\prime} ) & = \frac{a_{n} ( i \omega ) - a_{n} [ i ( \omega^{\prime} - n \omega_{d})]}{i ( \omega - \omega^{\prime} + n \omega_{d} ) }.
	\end{aligned}
\end{equation}
Both $a_{n}$ and $F_{n}$ are functions of $t$, but we do not write the explicit dependence in an effort to keep the notation simple. Finally, we rewrite the mean energy of $\env_{\alpha}$ as
\begin{equation}
    \begin{aligned}
        \langle H_{\mathcal{E}_{\alpha}} \rangle & = \frac{1}{2} \, \text{tr} \left[ \omega_{\alpha} \, f_{\alpha} ( \omega_{\alpha} ) \right] \\
        & + \frac{1}{4} \int_{0}^{\infty} d \omega \int_{0}^{t} dt^{\prime} \int_{0}^{t} dt^{\prime \prime} \, \omega \, \text{cos} [ \omega (t^{\prime} - t^{\prime \prime}) ]  \text{tr} \left\{ I_{\alpha} (\omega) \, \langle \{ X^{h} (t^{\prime}) , X^{h T} (t^{\prime \prime}) \} \rangle \right\} \\
        & - \frac{1}{4} \int_{0}^{\infty} d \omega \int_{0}^{t} dt^{\prime} \int_{0}^{t} dt^{\prime \prime} \, \omega \, \text{sin} [ \omega (t^{\prime} - t^{\prime \prime}) ]  \text{tr} \left\{ I_{\alpha} (\omega) \, \langle i \, [X^{h} (t^{\prime}) , X^{h T} (t^{\prime \prime})] \rangle \right\} f_{\alpha} ( \omega ) \\
        & + \frac{1}{2} \, \text{tr} \left\{ m_{\alpha}^{-1} \, g_{\alpha} ( \omega_{\alpha} ) \, \text{Im} \left[ \int_{0}^{t} dt^{\prime} \, e^{ i \omega_{\alpha} t^{\prime}} C_{\alpha}^{T} \int_{0}^{t^{\prime}} dt^{\prime \prime} \, G (t^{\prime},t^{\prime \prime}) \, C_{\alpha} \, e^{i \omega_{\alpha} t^{\prime \prime}} \right] \right\} \\
        & + \frac{1}{2} \, \text{tr} \left\{ m_{\alpha}^{-1} \, D_{\alpha} ( \omega_{\alpha} ) \, \text{Re} \left[ \int_{0}^{t} dt^{\prime} \, e^{ i \omega_{\alpha} t^{\prime}} C_{\alpha}^{T} \int_{0}^{t^{\prime}} dt^{\prime \prime} \, G (t^{\prime},t^{\prime \prime}) \, C_{\alpha} \, e^{i \omega_{\alpha} t^{\prime \prime}} \right] \right\} \\
        & + \frac{1}{8} \sum_{\beta} \int_{0}^{\infty} d \omega^{\prime} \int_{0}^{\infty} d \omega \, \omega^{\prime} \, \text{tr} \left[ I_{\alpha} (\omega^{\prime}) \, \mathcal{K} (\omega , -\omega^{\prime} , t) \, I_{\beta} ( \omega ) \, \mathcal{K}^{\dagger} (\omega , -\omega^{\prime} , t) \right] \, [ f_{\beta} ( \omega ) + f_{\alpha} ( \omega^{\prime} ) ] \\
        & + \frac{1}{8} \sum_{\beta} \int_{0}^{\infty} d \omega^{\prime} \int_{0}^{\infty} d \omega \, \omega^{\prime} \, \text{tr} \left[ I_{\alpha} (\omega^{\prime}) \, \mathcal{K} (\omega , \omega^{\prime} , t) \, I_{\beta} ( \omega ) \, \mathcal{K}^{\dagger} (\omega , \omega^{\prime} , t) \right] \, [ f_{\beta} ( \omega ) - f_{\alpha} ( \omega^{\prime} ) ] \\
        & + \frac{1}{4} \sum_{\beta} \text{Re} \left\{ \int_{0}^{\infty} d \omega^{\prime} \int_{0}^{\infty} d \omega \, \omega^{\prime} \, \text{tr} \left[ I_{\alpha} (\omega^{\prime}) \, \mathcal{K} (\omega , -\omega^{\prime} , t) \, I_{\beta} ( \omega ) \, \mathcal{K}^{T} (\omega , \omega^{\prime} , t) \right] \, g_{\alpha} ( \omega ) \right\} \\
        & + \frac{1}{4} \sum_{\beta} \text{Im} \left\{ \int_{0}^{\infty} d \omega^{\prime} \int_{0}^{\infty} d \omega \, \omega^{\prime} \, \text{tr} \left[ I_{\alpha} (\omega^{\prime}) \, \mathcal{K} (\omega , -\omega^{\prime} , t) \, I_{\beta} ( \omega ) \, \mathcal{K}^{T} (\omega , \omega^{\prime} , t) \right] \, D_{\alpha} ( \omega ) \right\}
    \end{aligned}
    \label{energy3}
\end{equation}
Equation \eqref{energy3} above is an exact expression valid for all times. Below we will study its long-time limit and compute the heat currents from it.

\subsection{Long-time behavior of the mean energy}
In this section we will study the long-time behavior of the mean energy of one thermal reservoir, which is exactly given by the Eq. \eqref{energy3}. Provided that a stable stationary regime for $\sys$ exists, the terms involving $X^{h}$ decay to zero for long times. Therefore, in this limit, the time dependence of $\langle H_{\mathcal{E}_{\alpha}} \rangle$ is contained in the $\mathcal{K}$ function alone (we will ignore the fourth and fifth terms on the right-hand side for now; see the note below). Moreover, we will study the average over a period $\tau_{d}$ of the time derivative of Eq. \eqref{energy3} and, for this reason, we can drop all constant terms appearing in such equation (among them, the contribution of the thermal energy and other terms, which we will show below). To complete our calculation and analyze the long-time limit of the energy, we need to examine the last four terms in Eq. \eqref{energy3}. Here we discuss in detail the behavior of the second one of those terms (since the analysis of the others is completely analogous). Using the expression for $\mathcal{K}$ above, the trace in that term can be written as
\begin{equation}
    \begin{aligned}
        & \text{tr} [ I_{\alpha} (\omega^{\prime}) \mathcal{K} (\omega , \omega^{\prime} , t) I_{\beta} ( \omega ) \mathcal{K}^{\dagger} (\omega , \omega^{\prime} , t)] \\
        & = \sum_{n} t^{2} \, \text{sinc}^{2} [(\omega - \omega^{\prime} + n \omega_{d})t/2] \text{tr} [ I_{\alpha} (\omega^{\prime}) a_{n} (i\omega) I_{\beta} ( \omega ) a_{n}^{\dagger} (i\omega) ] \\
        & + \sum_{n,m \neq n} t^{2} \, \text{sinc} [(\omega - \omega^{\prime} + n \omega_{d}) t/2] \, \text{sinc} [(\omega - \omega^{\prime} + m \omega_{d})t/2] \, e^{i(n-m)\omega_{d}t/2} \, \text{tr} [ I_{\alpha} (\omega^{\prime}) a_{n} (i\omega) I_{\beta} ( \omega ) a_{m}^{\dagger} (i\omega)] \\
        & + \sum_{n,m} 2 \, t \, \text{sinc} [(\omega - \omega^{\prime} + n \omega_{d})t/2] \, \text{Re} \left\{ e^{i ( \omega - \omega^{\prime} + n \omega_{d})t/2} \, \text{tr} [ I_{\alpha} (\omega^{\prime}) a_{n} (i\omega) I_{\beta} ( \omega ) F_{m}^{\dagger} (\omega , \omega^{\prime})] \right\} \\
        & + \sum_{n,m} \text{tr} [ I_{\alpha} (\omega^{\prime}) F_{n} (\omega , \omega^{\prime}) I_{\beta} ( \omega ) F_{m}^{\dagger} (\omega , \omega^{\prime})].
    \end{aligned}
    \label{trace}
\end{equation}
As we mentioned above, our goal is to study the average over a period $\tau_{d}$ of the time derivative of $\langle H_{\env_{\alpha}} \rangle$. Therefore, we will only keep terms which are linear in time (the ones that oscillate or are constant will vanish after differentiation and averaging). In order to find those terms, we will make use of the fact that
\begin{equation}
    \lim_{t \to \infty} t \, \text{sinc} (\omega t) = \lim_{t \to \infty} t \, \text{sinc}^{2} (\omega t) = 2 \pi \, \delta (\omega),
\end{equation}
and \begin{equation}
    \lim_{t \to \infty} a_{n} (i \omega) = \tilde{A}_{n} (i \omega),
\end{equation}
where $\tilde {A}_{n} (i \omega)$ is to be understood as
\begin{equation}
    \tilde {A}_{n} (i \omega) = \lim_{\sigma \to 0^{+}} \tilde{A}_{n} (\sigma + i \omega).
\end{equation}
Also, in the long-time limit, $F_{n}$ is constant. Therefore, in the long-time limit, the second term on the right-hand side of Eq. \eqref{trace} oscillates with frequencies that are multiples of $\omega_{d}$, and the third and fourth one tend to a constant. Thus, all these terms can be disregarded and the only relevant one turns out to be the first one, which behaves as
\begin{equation}
    t^{2} \text{sinc}^{2} [(\omega - \omega^{\prime} + n \omega_{d})t/2] \text{tr} [ I_{\alpha} (\omega^{\prime}) a_{n} (\omega) I_{\beta} ( \omega ) a_{n}^{\dagger} (\omega) ] \to 2 \pi t \delta (\omega - \omega^{\prime} + n \omega_{d}) \text{tr} [ I_{\alpha} (\omega^{\prime}) \tilde{A}_{n} (i \omega) I_{\beta} ( \omega ) \tilde{A}_{n}^{\dagger} (i \omega) ].
\end{equation}
Hence, for long times, the relevant terms of the mean energy of $\env_{\alpha}$ are
\begin{equation}
    \begin{aligned}
        \langle H_{\mathcal{E}_{\alpha}} \rangle & \to \frac{\pi}{4} \sum_{n,\beta} \int_{0}^{\infty} d \omega \, \omega \, \text{tr} [ I_{\alpha} (\omega)  \tilde{A}_{n} [ i ( \omega - n \omega_{d} ) ]  I_{\beta} (\omega - n \omega_{d}) \tilde{A}_{n}^{\dagger} [ i ( \omega - n \omega_{d} ) ] ] [ f_{\beta} (\omega - n \omega_{d}) - f_{\alpha} ( \omega ) ] t
    \end{aligned}
    \label{energy4}
\end{equation}
where the summation is over all $n \in \mathbb{Z}$ and we used the property $\tilde{A}_{n} ( i \omega) = \tilde{A}_{-n}^{\ast} (- i \omega)$. All other terms are either constant or oscillatory and they vanish when averaged for long times.

Note regarding the terms in Eq. \eqref{energy3} that include the functions $g_{\alpha}$ and $D_{\alpha}$. These terms appear to significantly contribute in the long-time average only in the case in which $\omega_{a} = k \omega_{d} / 2$ with $k$ some integer but, if this were true, it would create a discontinuity in the average value of the heat current (as a function of $\omega_{\alpha}$). This, of course, cannot be, because $\langle H_{\mathcal{E}_{\alpha}} \rangle$ and its derivative are continuous functions of $\omega_{\alpha}$ for any $t>0$ and, thus, their average for long times must be a continuous function too. Therefore, their contribution must vanish (due to a condition analogous to Eq. \eqref{unitary} that, so far, remains elusive) and the correct expression for the case $\omega_{a} = k \omega_{d} / 2$ is the one we obtained above (Eq. \eqref{energy4}), which is continuous for any value of $\omega_{\alpha}$.

\subsection{Average heat current of a reservoir}

\begin{figure}[htp]
    \begin{center}
    \includegraphics[scale=2]{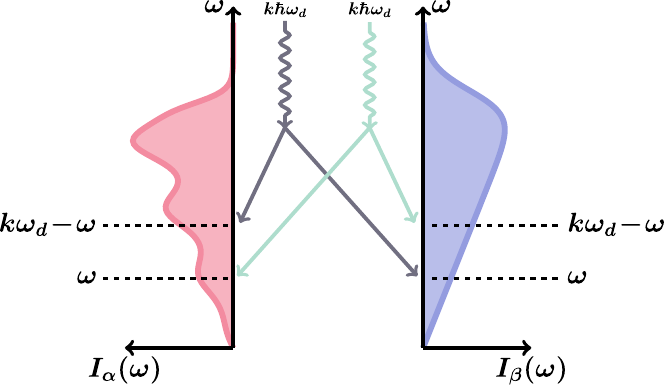}
    \caption{Nonresonant processes between two reservoirs $\env_{\alpha}$ and $\env_{\beta}$.}
    \label{fig:NR}
\end{center}
\end{figure}

\begin{figure}[htp]
    \begin{center}
    \includegraphics[scale=2]{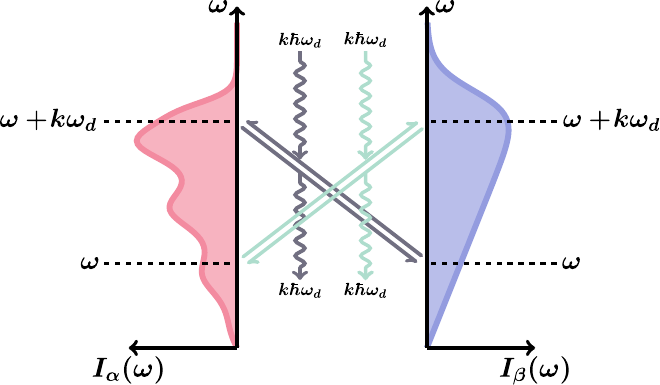}
    \caption{Resonant processes between two reservoirs $\env_{\alpha}$ and $\env_{\beta}$.}
    \label{fig:R}
\end{center}
\end{figure}

In this section we will compute the average heat current exchanged with a thermal reservoir in the long-time limit, $\dot{\bar{Q}}_{\alpha}$, which, as we previously said, is obtained by averaging $d\langle H_{\mathcal{E}_{\alpha}} \rangle/dt$ in a period of the driving for long times. Then, we will show that the obtained expression can be split as the sum of a contribution arising from resonant and another one arising from nonresonant processes. From Eq. \eqref{energy4} it is immediate that $\dot{\bar{Q}}_{\alpha}$ is
\begin{equation}
    \dot{\bar{Q}}_{\alpha} = \frac{\pi}{4} \sum_{n,\beta} \int_{0}^{\infty} d \omega \, \omega \, \text{tr} [ I_{\alpha} (\omega)  \tilde{A}_{n} [ i ( \omega - n \omega_{d} ) ]  I_{\beta} (\omega - n \omega_{d}) \tilde{A}_{n}^{\dagger} [ i ( \omega - n \omega_{d} ) ] ]  [ f_{\beta} (\omega - n \omega_{d}) - f_{\alpha} ( \omega) ].
    \label{q1}
\end{equation}
In order to separate the contributions of the resonant and nonresonant processes, the first step is to split the summation for positive and negative values of $n$:
\begin{equation}
    \begin{aligned}
        \dot{\bar{Q}}_{\alpha} & = \frac{\pi}{4} \sum_{\beta} \int_{0}^{\infty} d \omega \, \omega \, \text{tr} [ I_{\alpha} (\omega)  \tilde{A}_{0} ( i \omega )  I_{\beta} (\omega) \tilde{A}_{0}^{\dagger} (i \omega) ]  [ f_{\beta} (\omega) - f_{\alpha} ( \omega ) ] \\
        & + \frac{\pi}{4} \sum_{n>0,\beta} \int_{0}^{\infty} d \omega \, \omega \, \text{tr} [ I_{\alpha} (\omega)  \tilde{A}_{n} [ i ( \omega - n \omega_{d} ) ]  I_{\beta} (\omega - n \omega_{d}) \tilde{A}_{n}^{\dagger} [ i ( \omega - n \omega_{d} ) ] ] [ f_{\beta} (\omega - n \omega_{d}) - f_{\alpha} ( \omega ) ] \\
        & + \frac{\pi}{4} \sum_{n<0,\beta} \int_{0}^{\infty} d \omega \, \omega \, \text{tr} [ I_{\alpha} (\omega)  \tilde{A}_{n} [ i ( \omega - n \omega_{d} ) ]  I_{\beta} (\omega - n \omega_{d}) \tilde{A}_{n}^{\dagger} [ i ( \omega - n \omega_{d} ) ] ]  [ f_{\beta} (\omega - n \omega_{d}) - f_{\alpha} ( \omega ) ].
    \end{aligned}
    \label{q2}
\end{equation}
The first term on the right hand side of Eq. \eqref{q2}, which we will call $\dot{\bar{Q}}_{\alpha}^{ST}$, corresponds to the static heat current and it is such that $\sum_{\alpha} \dot{\bar{Q}}_{\alpha}^{ST} = 0$. That is, there is no net heat flow in the absence of driving. With a change of variables in the integrand we can write the second term on the right hand side of Eq. \eqref{q2} as
\begin{equation}
    \begin{aligned}
        & \sum_{n>0,\beta} \int_{0}^{\infty} d \omega \, \omega \, \text{tr} [ I_{\alpha} (\omega)  \tilde{A}_{n} [ i ( \omega - n \omega_{d} ) ]  I_{\beta} (\omega - n \omega_{d}) \tilde{A}_{n}^{\dagger} [ i ( \omega - n \omega_{d} ) ] ]  \{ f_{\beta} (\omega - n \omega_{d}) - f_{\alpha} ( \omega ) \} \\
        & = \sum_{k>0,\beta} \int_{0}^{k \omega_{d}} d \omega \, \omega \, \text{tr} [ I_{\alpha} (\omega) \tilde{A}_{k} [ i ( \omega - k \omega_{d} ) ] I_{\beta} (k \omega_{d} - \omega) \tilde{A}_{k}^{\dagger} [ i ( \omega - k \omega_{d} ) ] ] [ f_{\beta} ( k \omega_{d} - \omega ) + f_{\alpha} ( \omega ) ] \\
        & + \sum_{k>0,\beta} \int_{0}^{\infty} d \omega \, (\omega + k \omega_{d}) \, \text{tr} [ I_{\alpha} (\omega + k \omega_{d}) \tilde{A}_{k} (i\omega) I_{\beta} (\omega) \tilde{A}_{k}^{\dagger} (i\omega) ] [ f_{\beta} ( \omega ) - f_{\alpha} (\omega + k\omega_{d}) ].
    \end{aligned}
\end{equation}
Lastly, on the third term on the right-hand side of Eq. \eqref{q2} we change the sign of the $n$'s:
\begin{equation}
    \begin{aligned}
        & \sum_{n<0,\beta} \int_{0}^{\infty} d \omega \, \omega \, \text{tr} [ I_{\alpha} (\omega)  \tilde{A}_{n} [ i ( \omega - n \omega_{d} ) ]  I_{\beta} (\omega - n \omega_{d}) \tilde{A}_{n}^{\dagger} [ i ( \omega - n \omega_{d} ) ] ]  [ f_{\beta} ( \omega - n \omega_{d} ) - f_{\alpha} ( \omega ) ] \\
        & = \sum_{k>0,\beta} \int_{0}^{\infty} d \omega \, \omega \, \text{tr} [ I_{\alpha} (\omega)  \tilde{A}_{-k} [ i ( \omega + k \omega_{d} ) ]  I_{\beta} (\omega + k \omega_{d}) \tilde{A}_{-k}^{\dagger} [ i ( \omega + k \omega_{d} ) ] ]  [ f_{\beta}  ( \omega + k \omega_{d} ) - f_{\alpha} ( \omega ) ].
    \end{aligned}
\end{equation}
Note that in all summations we replaced the index $n$ that was an integer for $k$, that can only be a positive integer to be consistent with the equations in main text. Introducing the changes described above, Eq. \eqref{q2} can be written as
\begin{equation}
    \begin{aligned}
        \dot{\bar{Q}}_{\alpha} & = \frac{\pi}{4} \sum_{\beta} \int_{0}^{\infty} d \omega \, \omega \, \text{tr} [ I_{\alpha} (\omega)  \tilde{A}_{0} ( i \omega )  I_{\beta} (\omega) \tilde{A}_{0}^{\dagger} (i \omega) ]  [ f_{\beta} ( \omega ) - f_{\alpha} ( \omega ) ] \\
        & + \frac{\pi}{4} \sum_{k>0,\beta} \int_{0}^{k \omega_{d}} d \omega \, \omega \, \text{tr} [ I_{\alpha} (\omega) \tilde{A}_{k} [ i ( \omega - k \omega_{d} ) ] I_{\beta} (k \omega_{d} - \omega) \tilde{A}_{k}^{\dagger} [ i ( \omega - k \omega_{d} ) ] ] [ f_{\beta} ( k \omega_{d} - \omega ) + f_{\alpha} ( \omega ) ] \\
        & + \frac{\pi}{4} \sum_{k>0,\beta} \int_{0}^{\infty} d \omega \, (\omega + k \omega_{d}) \, \text{tr} [ I_{\alpha} (\omega + k \omega_{d}) \tilde{A}_{k} (i\omega) I_{\beta} (\omega) \tilde{A}_{k}^{\dagger} (i\omega) ] [ f_{\beta} ( \omega ) - f_{\alpha} (\omega + k\omega_{d}) ] \\
        & + \frac{\pi}{4} \sum_{k>0,\beta} \int_{0}^{\infty} d \omega \, \omega \, \text{tr} [ I_{\alpha} (\omega)  \tilde{A}_{-k} [ i ( \omega + k \omega_{d} ) ]  I_{\beta} (\omega + k \omega_{d}) \tilde{A}_{-k}^{\dagger} [ i ( \omega + k \omega_{d} ) ] ]  [ f_{\beta} ( \omega + k \omega_{d} ) - f_{\alpha} ( \omega ) ].
    \end{aligned}
    \label{q3}
\end{equation}
Thus, we can write the total heat current exchanged with $\env_{\alpha}$ as
\begin{equation}
    \dot{\bar{Q}}_{\alpha} = \dot{\bar{Q}}_{\alpha}^{ST} + \dot{\bar{Q}}_{\alpha}^{NR} + \dot{\bar{Q}}_{\alpha}^{R}.
\end{equation}
The contribution of the nonresonant processes, $\dot{\bar{Q}}_{\alpha}^{NR}$, is contained in the second term on the right-hand side of Eq. \eqref{q3} and can be written as
\begin{equation}
    \dot{\bar{Q}}_{\alpha}^{NR} = \sum_{k>0,\beta} \int_{0}^{k \omega_{d}} d \omega \, (k \omega_{d} - \omega) \, \tilde{p}_{\alpha,\beta}^{(k)} (\omega) [n_{\beta} (\omega) + 1/2] + \sum_{k>0,\beta} \int_{0}^{k \omega_{d}} d \omega \, \omega \, \tilde{p}_{\beta,\alpha}^{(k)} (\omega) [n_{\alpha} (\omega) + 1/2],
    \label{qNRAp}
\end{equation}
where in the first term we made a change of variable in the integrand and in the second one, we used the translation of the argument property of the $\tilde{A}_{k}$ coefficients. Also, we defined the emission rate per unit frequency
\begin{equation}
    \tilde{p}_{\alpha,\beta}^{(k)} (\omega) = \frac{\pi}{2} \text{tr} [ I_{\alpha} (k \omega_{d} - \omega) \tilde{A}_{-k}  (i \omega) I_{\beta} (\omega) \tilde{A}_{-k}^{\dagger} (i \omega ) ]
\end{equation}
and used the relation $f_{\alpha} ( \omega ) = 2 n_{\alpha} (\omega) + 1$. The nonresonant processes correspond to the creation of two pairs of excitations in the reservoirs, one in the mode of frequency $k\omega_{d} - \omega$ and the other one, in the one with frequency $\omega$, such that both frequencies add up to $k \omega_{d}$. The first term on the right-hand side of Eq. \eqref{qNR} corresponds to the case in which mode $k\omega_{d} - \omega$ is excited in $\env_{\alpha}$, while mode $\omega$ is excited in $\env_{\beta}$. Analogously, the second term corresponds to the opposite case: mode $\omega$ is excited in $\env_{\alpha}$ and mode $k\omega_{d} - \omega$ is excited in $\env_{\beta}$. These processes are depicted in Fig. \ref{fig:NR}. The first term on the right hand side of Eq. \eqref{qNRAp} is represented in gray and the second one, in green.
On the other hand, the contribution of the resonant processes, $\dot{\bar{Q}}_{\alpha}^{R}$, comes from the third and fourth terms on the right-hand side of Eq. \eqref{q3}. These can be written as
\begin{equation}
    \begin{aligned}
        \dot{\bar{Q}}_{\alpha}^{R} & = \sum_{k>0,\beta} \int_{0}^{\infty} d \omega \, (\omega + k \omega_{d}) \, p_{\alpha,\beta}^{(k)} (\omega) n_{\beta} (\omega) - \sum_{k>0,\beta} \int_{0}^{\infty} d \omega \, (\omega + k \omega_{d}) \, p_{\beta,\alpha}^{(-k)} (\omega+k\omega_{d}) n_{\alpha} (\omega + k \omega_{d}) \\
        & + \sum_{k>0,\beta} \int_{0}^{\infty} d \omega \, \omega \, p_{\alpha,\beta}^{(-k)} (\omega+k\omega_{d}) n_{\beta} (\omega + k \omega_{d}) - \sum_{k>0,\beta} \int_{0}^{\infty} d \omega \, \omega \, p_{\beta,\alpha}^{(k)} (\omega) n_{\alpha} (\omega)
    \end{aligned}
    \label{qRAp}
\end{equation}
where we defined the transition rate per unit frequency
\begin{equation}
    p_{\alpha,\beta}^{(k)} (\omega) = \frac{\pi}{2} \text{tr} [ I_{\alpha} (\omega + k \omega_{d}) \tilde{A}_{k} (i\omega) I_{\beta} (\omega) \tilde{A}_{k}^{\dagger} (i\omega) ],
\end{equation}
and used the translation of the argument property to arrive at the second and third terms on the right-hand side of Eq. \eqref{qRAp}. The resonant processes correspond to the transport of excitations between different environmental modes, due to absorption (or emission) of energy from (or into) the driving field. The first two terms on the right-hand side of Eq. \eqref{qRAp} correspond to opposite processes. The first one, being positive, indicates that an excitation is being transported from the mode $\omega$ in $\env_{\beta}$ to the mode $\omega + k \omega_{d}$ in $\env_{\alpha}$, while absorbing a packet of energy $k \omega_{d}$ from the driving field. The second one corresponds to the reverse process: an excitation is being transported from the mode $\omega + k \omega_{d}$ in $\env_{\alpha}$ to the mode $\omega$ in $\env_{\beta}$, while dumping $k \omega_{d}$ into the driving field. Analogously, the third and fourth terms on the right-hand side of Eq. \eqref{qRAp} correspond to opposite processes too. The third one describes an excitation being transported from the mode $\omega + k \omega_{d}$ in $\env_{\beta}$ to a mode $\omega$ in $\env_{\alpha}$, while dumping $k \omega_{d}$ into the driving field. The fourth one, in turn, describes an excitation being transported from the mode $\omega$ in $\env_{\alpha}$ to a mode $\omega + k \omega_{d}$ in $\env_{\beta}$, while absorbing $k \omega_{d}$ from the driving field. These processes are depicted in Fig. \ref{fig:R}. The first two terms on the right hand side of Eq. \eqref{qRAp} are represented in gray, and the third and fourth ones, in green.

Notably, using the translation of the argument property of the $\tilde{A}_{k}$ coefficients, one can show that the transition rates per unit frequency of opposite processes are equal. That is,
\begin{equation}
    p_{\beta,\alpha}^{(-k)} ( \omega + k \omega_{d}) = p_{\alpha,\beta}^{(k)} (\omega).
\end{equation}
This is nothing but the principle of detailed balance. Using this relation, Eq. \eqref{qRAp} can be written as the sum of two terms only:
\begin{equation}
    \dot{\bar{Q}}_{\alpha}^{R} = \sum_{k>0,\beta} \int_{0}^{\infty } d \omega \, (\omega + k \omega_{d}) \, p_{\alpha,\beta}^{(k)} (\omega) [n_{\beta} (\omega) - n_{\alpha} (\omega + k \omega_{d})] - \sum_{k>0,\beta} \int_{0}^{\infty} d \omega \, \omega \, p_{\beta,\alpha}^{(k)} (\omega) [n_{\alpha} (\omega) - n_{\beta} (\omega + k \omega_{d})].
    \label{qR2}
\end{equation}
Equation \eqref{qR2} shows that, although the transition rate is the same for opposite processes, there is a net flow of excitations in the direction in which the occupation number is lowest. For example, if $n_{\beta} (\omega) > n_{\alpha} (\omega + k \omega_{d})$ in the first term on the right hand side of Eq. \eqref{qR2}, then the net transport of excitations is from the mode $\omega$ in $\env_{\beta}$ to the mode $\omega + k \omega_{d}$ in $\env_{\alpha}$.

\subsection{Total heat current}

The total heat exchanged $\Delta \mathcal{Q}$ is obtained from the equations above by adding the contributions from all reservoirs. That is, $\Delta \mathcal{Q} = \tau_{d} \sum_{\alpha} \dot{\bar{Q}}_{\alpha}$. The only computation needed is a change of indices in the summation ($\alpha \leftrightarrow \beta$) to cancel duplicate terms. Explicitly, we have
\begin{equation}
    \Delta \mathcal{Q}^{NR} = \tau_{d} \sum_{k,\alpha,\beta} k \omega_{d} \int_{0}^{k \omega_{d}} d \omega \, \tilde{p}_{\alpha,\beta}^{(k)} (\omega) [n_{\beta} (\omega) + 1/2]
\end{equation}
and
\begin{equation}
    \Delta \mathcal{Q}^{R} = \tau_{d} \sum_{k,\alpha,\beta} k \omega_{d} \int_{0}^{\infty } d \omega \, p_{\alpha,\beta}^{(k)} (\omega) [n_{\beta} (\omega) - n_{\alpha} (\omega + k \omega_{d})]
\end{equation}
where the summation is over positive integer $k$'s. This proves Eqs. \eqref{qNR} and \eqref{qR} of the main text, which are the basis of our analysis for the properties of linear thermal engines.

\section{Relation between $\dot{\bar{Q}}_{\sys}$ and $\dot{\bar{Q}}_{\env}$ \label{sec:heatCurrents}}

In this section we will prove the statement given in the Sec. \ref{sec:workHeat} in the main text: the average heat current $\dot{\bar{Q}}_{\sys}$ is determined by the average variation of energy of the reservoirs. In order to do this, we first consider the definitions of the heat currents of the working medium and reservoirs, respectively:
\begin{equation}
	\begin{aligned}
		\dot{Q}_{\sys} & = - i \langle [ H_{\sys} , H_{\sys, \env} ] \rangle \\
		\dot{Q}_{\env} & = - i \langle [ H_{\env}, H_{\sys, \env} ] \rangle.
	\end{aligned}
\end{equation}
It is evident that $\dot{Q}_{\sys} + \dot{Q}_{\env} = - i \langle [ H , H_{\sys, \env} ] \rangle  = - d \langle H_{\sys, \env} \rangle / dt$. We will now show that $d \langle H_{\sys, \env} \rangle / dt$ can be written as the derivative of a periodic function and, thus, when averaged over a period of the driving for long times it vanishes. Therefore, we obtain $\dot{\bar{Q}}_{\sys} + \dot{\bar{Q}}_{\env} = 0$, as we stated in the main text. A straightforward computation shows that
\begin{equation}
	\begin{aligned}
		\dot{Q}_{\sys} & = - \langle P^{T} M^{-1} \sum_{\alpha} C_{\alpha} q_{\alpha} \rangle \\
		\dot{Q}_{\env} & = \langle X^{T} \sum_{\alpha} C_{\alpha} m_{\alpha}^{-1} p_{\alpha} \rangle.
	\end{aligned}
\end{equation}
Finally, using Heisenberg's equation of motion to eliminate the coordinates of the environment we arrive at
\begin{equation}
	\dot{Q}_{\sys} + \dot{Q}_{\env} = \frac{d}{dt} \langle X^{T} [ \dot{P} + V(t) X] \rangle.
\end{equation}
As the stationary state of $\sys$ is $\tau_{d}$ periodic, the right-hand side of the previous equation is the derivative of a $\tau_{d}$-periodic function, as we wanted to prove.

Note that this implies that the it is not necessary to include the interaction energy when defining the work produced and heat exchanged by the engine in the main text in Eq. \eqref{variationHS} (as it is customary in strong-coupling thermodynamics \cite{talknerHanggi,espositoLindenbergVDB,katoTanimura,strasbergSchallerBrandes}), because it will vanish when computing the average value of its derivative.

\section{Clausius inequality \label{sec:clausiusAp}}

In this section we will derive the form of Clausius inequality shown in Eq. \eqref{clausius} in the main text. We begin from the equation for the heat flowing from the working medium to the reservoir:
\begin{equation}
	\begin{aligned}
	\Delta \mathcal{Q}_{\env \leftarrow \sys}^{R} / \tau_{d} & = - \sum_{k,\alpha,\beta} \int_{\mathcal{I}^{-}} d \omega \, \omega \, p_{\alpha \beta}^{(k)} (\omega) [ n_{\beta} (\omega) - n_{\alpha} (\omega + k \omega_{d}) ] \\
	& + \sum_{k,\alpha,\beta} \int_{\mathcal{I}^{+}} d \omega (\omega + k \omega_{d}) p_{\alpha \beta}^{(k)} (\omega) [ n_{\beta} (\omega) - n_{\alpha} (\omega + k \omega_{d}) ]
	\end{aligned}
	\label{clausius1}
\end{equation}
In the first term in Eq. \eqref{clausius1} we have $n_{\beta} (\omega) < n_{\alpha} (\omega + k \omega_{d})$ and, therefore, $\omega > [\Omega_{\beta} (\omega) / \Omega_{\alpha} (\omega + k \omega_{d})] (\omega + k \omega_{d})$. Conversely, in the second term we have $n_{\beta} (\omega) > n_{\alpha} (\omega + k \omega_{d})$ and, therefore, $\omega + k \omega_{d} > [\Omega_{\alpha} (\omega + k \omega_{d}) / \Omega_{\beta} (\omega)] \omega$. By replacing these inequalities in the integrands of their respective terms, and considering the minimum of the quotient of characteristic frequencies in order to write them outside the integrals and summations, we arrive at
\begin{equation}
	\begin{aligned}
	\Delta \mathcal{Q}_{\env \leftarrow \sys}^{R} / \tau_{d} & > - \text{min} \{ \Omega_{\beta} (\omega) / \Omega_{\alpha} (\omega + k \omega_{d}) \} \sum_{k,\alpha,\beta} \int_{\mathcal{I}^{-}} d \omega (\omega + k \omega_{d}) p_{\alpha \beta}^{(k)} (\omega) [ n_{\beta} (\omega) - n_{\alpha} (\omega + k \omega_{d}) ] \\
	& + \text{min} \{ \Omega_{\alpha} (\omega + k \omega_{d}) / \Omega_{\beta} (\omega) \} \sum_{k,\alpha,\beta} \int_{\mathcal{I}^{+}} d \omega \, \omega \, p_{\alpha \beta}^{(k)} (\omega) [ n_{\beta} (\omega) - n_{\alpha} (\omega + k \omega_{d}) ].
	\end{aligned}
	\label{clausius2}
\end{equation}
Equation \eqref{clausius2} can be bounded again by considering the minimum between both minimums:
\begin{equation}
	\begin{aligned}
	\Delta \mathcal{Q}_{\env \leftarrow \sys}^{R} / \tau_{d} & > \text{min} \{ m , 1/M \} \left\{ - \sum_{k,\alpha,\beta} \int_{\mathcal{I}^{-}} d \omega (\omega + k \omega_{d}) p_{\alpha \beta}^{(k)} (\omega) [ n_{\beta} (\omega) - n_{\alpha} (\omega + k \omega_{d}) ] \right. \\
	& \left. + \sum_{k,\alpha,\beta} \int_{\mathcal{I}^{+}} d \omega \, \omega \, p_{\alpha \beta}^{(k)} (\omega) [ n_{\beta} (\omega) - n_{\alpha} (\omega + k \omega_{d}) ] \right\},
	\end{aligned}
	\label{clausius3}
\end{equation}
where $m = \text{min} \{ \Omega_{\alpha} (\omega) / \Omega_{\beta} (\omega + k \omega_{d}) \}$ and $M = \text{max} \{ \Omega_{\alpha} (\omega) / \Omega_{\beta} (\omega + k \omega_{d}) \}$ and we used the fact that, for any function $f$, $\text{min} \{ 1 / f \} = 1 / \text{max} \{ f \}$. By noticing that the right hand side of Eq. \eqref{clausius3} is $\lvert \Delta \mathcal{Q}_{\env \rightarrow \sys}^{R} \rvert / \tau_{d} $, we arrive at the desired inequality.

\section{Cost of preparing nonthermal reservoirs \label{sec:costAp}}

Consider the definition of the efficiency of the engine shown in the main text (Eq. \eqref{eff}):
\begin{equation}
	\eta = \frac{\lvert \Delta \mathcal{Q}_{\env \rightarrow \sys}^{R} \rvert - \Delta \mathcal{Q}_{\env \leftarrow \sys}^{R}-\Delta \mathcal{Q}^{NR}}{\lvert \Delta \mathcal{Q}_{\env \rightarrow \sys}^{R} \rvert}.
\end{equation}
This would be the efficiency of the engine computed in step (iii) of the process described there. Now, let us include the energy invested $\mathfrak{C}$ in step (ii) to transform a thermal reservoir in a nonthermal one. Since this cost does not reduce the amount of work done by the engine, it should be included only in the denominator of $\eta$ (which represents the total energy invested). We call this new efficiency $\tilde{\eta}$:
\begin{equation}
	\tilde{\eta} = \frac{\lvert \Delta \mathcal{Q}_{\env \rightarrow \sys}^{R} \rvert - \Delta \mathcal{Q}_{\env \leftarrow \sys}^{R}-\Delta \mathcal{Q}^{NR}}{\lvert \Delta \mathcal{Q}_{\env \rightarrow \sys}^{R} \rvert + \mathfrak{C}} = \frac{1}{1 + \mathfrak{C} / \lvert \Delta \mathcal{Q}_{\env \rightarrow \sys}^{R} \rvert} \eta.
\end{equation}
In the main text we showed that $\eta < \eta_{g}$, thus
\begin{equation}
	\tilde{\eta} < \frac{1}{1 + \mathfrak{C} / \lvert \Delta \mathcal{Q}_{\env \rightarrow \sys}^{R} \rvert} \eta_{g}.
\end{equation}
As we added the cost to obtain the nonthermal reservoir starting from a thermal one, this bound should be similar to the Carnot limit for the complete process:
\begin{equation}
	\frac{1}{1 + \mathfrak{C} / \lvert \Delta \mathcal{Q}_{\env \rightarrow \sys}^{R} \rvert} \eta_{g} \simeq \eta_{c}.
\end{equation}
Solving this equation for $\mathfrak{C} / \lvert \Delta \mathcal{Q}_{\env \rightarrow \sys}^{R} \rvert$, we obtain Eq. \eqref{cost} shown in the main text.

\twocolumngrid

\end{document}